\documentclass[%
 reprint,
 amsmath,amssymb,
 aps,
]{revtex4-2}

\usepackage{graphicx}
\usepackage{dcolumn}
\usepackage{bm}
\usepackage{hyperref}
\usepackage[dvipsnames]{xcolor}
\hypersetup{
    colorlinks,
    linkcolor={MidnightBlue!65!black},
    citecolor={MidnightBlue!70!black},
    urlcolor={MidnightBlue!70!black}
}
\usepackage{comment}
\usepackage{amsmath}

\newcommand{\cs}[1]{\textcolor{ForestGreen}{CS: #1}}

\begin{document}

\preprint{APS/123-QED}

\title{Inverse Galaxy-Galaxy Lensing: \\Magnification, Intrinsic Alignments and Cosmology}

\author{Dane N.~Cross}
\email{dcross@ifae.es}
\affiliation{\vspace{1mm}Departament de F\'{\i}sica, Universitat Aut\`{o}noma de Barcelona (UAB), 08193 Bellaterra (Barcelona), Spain}
\affiliation{Institut de F\'{\i}sica d'Altes Energies (IFAE), The Barcelona Institute of Science and Technology, Campus UAB, 08193 Bellaterra (Barcelona), Spain}
\author{Carles S{\'a}nchez} 
\email{carles.sanchez@uab.cat}
\affiliation{\vspace{1mm}Departament de F\'{\i}sica, Universitat Aut\`{o}noma de Barcelona (UAB), 08193 Bellaterra (Barcelona), Spain}
\affiliation{Institut de F\'{\i}sica d'Altes Energies (IFAE), The Barcelona Institute of Science and Technology, Campus UAB, 08193 Bellaterra (Barcelona), Spain}

\date{\today}

\begin{abstract}
Current and upcoming imaging galaxy surveys are pushing galaxy samples to higher and higher redshifts. This push will be more pronounced for lens galaxies, for which we only need to measure galaxy positions, not shapes. As a result, we will increasingly often have lens galaxy samples at redshifts higher than those of source galaxies, changing the traditional configuration of galaxy-galaxy lensing (GGL). In this paper, we explore this situation, where lens galaxies are behind source galaxies, which we call inverse galaxy-galaxy lensing (IGGL). We take projected lens and source sample specifications from the Vera Rubin Observatory LSST Dark Energy Science Collaboration (DESC) to compare astrophysical and cosmological constraints between traditional GGL and IGGL. We find IGGL to behave in a different way than GGL, being especially sensitive to lensing magnification, intrinsic alignments (IA) and cosmology, but largely independent of galaxy bias (as opposed to traditional GGL). In this way, we find IGGL can provide independent and robust cosmological constraints without combination with galaxy clustering, and can also probe IA at high redshift and baryonic effects at small scales without being entwined with the effects of non-linear galaxy bias. When combined with cosmic shear, we find IGGL to improve $S_8$ constraints by 25\% compared to cosmic shear alone, while also providing tighter and more robust constraints on IA and baryons. 
\end{abstract}

\maketitle

\section{Introduction}
\label{sec:intro}

Gravitational lensing is the effect of light rays being deflected when they propagate through an inhomogeneous gravitational field. As a consequence, the light of distant galaxies gets perturbed when passing close to accumulations of foreground mass, distorting our observed image of those galaxies in both shape and size, due to the effect of shear and magnification, respectively. The shear produced on galaxy shapes is tangential to the direction toward the center of the mass distribution causing the lensing, and the level of distortion on shapes and size is given by the properties of the matter distribution causing it (see \cite{2001PhR...340..291B} for a review). 

Galaxy-galaxy lensing (GGL, \citep{1992ApJ...388..272K}) refers to the correlation between (\textit{lens}) galaxy positions and (\textit{source}) galaxy shapes. Note that we will use this notation throughout the paper: \textit{lens} galaxies are those for which we only measure positions, and \textit{source} galaxies are those for which we measure positions and shapes, regardless of their redshifts. To make this more clear, we will sometimes refer to the samples involved as lens (position) sample and source (shape) sample.  To measure GGL, one averages the tangential component of the source shape ellipticity with respect to the lens-source direction over many pairs of lens and source galaxies, although there exist variations to this method \citep{Prat_2023}. Traditionally, galaxy positions are those of foreground lens galaxies, which trace the matter field causing the shape distortion (\textit{shear}) on more distant background source galaxies. When using galaxy samples to measure this effect in this traditional configuration, the redshift distribution of lens galaxies is at lower redshift than that of source galaxies. Measurements of GGL in this traditional setting have been improving in significance for more than two decades \citep{Brainerd1996,Sheldon2004,Mandelbaum_2005,Mandelbaum2010,Mandelbaum_2014,Clampitt2016,Prat_2019}, reaching a level of maturity which now yields signal-to-noise ratios of more than 100 in some cases \citep{Prat_2022,Zacharegkas_2021}.

Because lens galaxy positions are not perfect tracers of the matter distribution, galaxy-galaxy lensing in this traditional configuration is subject to the details of the galaxy-matter connection, in particular to the so-called galaxy bias (see \citep{Desjacques_2018} for a review). In this scenario, the dependence of GGL on the galaxy-matter connection is such that cosmological information (like the amplitude of matter fluctuations, $\sigma_8$) is degenerate with galaxy bias. For this reason, in order to use GGL to obtain cosmological constraints, it typically needs to be combined with other measurements, such as galaxy clustering or CMB lensing, to break the degeneracies between cosmological information and galaxy bias \citep{Kwan2016, Baxter2016, VanUitert2017, Joudaki2017, Singh_2019, Pandey_2022, y3-2x2ptaltlensresults}.  

The fact that GGL needs to be combined with other probes in order to constrain cosmology can in some cases pose challenges. The measurement of GGL itself, being a cross-correlation between galaxy positions and shapes, has been shown to be free of many systematic effects impacting the autocorrelation of positions (galaxy clustering) or shapes (cosmic shear) \citep{Clampitt2016,Y1GGL,Prat_2022}. For example, the measurement of galaxy clustering, which is traditionally combined with GGL to probe cosmology, is strongly affected by systematic effects such as galaxy density variations correlated with observing conditions (see \citep{Noah_weights} and references therein). In addition, the different weighting kernels in galaxy clustering and GGL can introduce de-correlations that can directly bias cosmological constraints in the presence of evolving astrophysical effects such as galaxy biasing \citep{Pandey2023}. Furthermore, it is difficult to extract valuable information from small scales in GGL and galaxy clustering, as they include contributions from both baryonic physics and non-linear galaxy bias, and these contributions are hard to disentangle \citep{mead2015, Harnois2015,HMCODE}. 

Beyond the dependence on cosmology, galaxy bias and baryonic effects, the galaxy-galaxy lensing measurement is also sensitive to the effects of intrinsic alignments (IA) and weak lensing magnification. On the one hand, the effect of IA comes from the alignment of the source galaxy orientations with the lens galaxy positions due to their physical association, not due to cosmological lensing \citep{Croft_2000,Heavens_2000,Mandelbaum_2006}. On the other hand, the weak lensing magnification of lens (position) galaxies is correlated with the tangential shear in the shapes of source galaxies \citep{Hildebrandt_2009,Van_Waerbeke_2010}. The modeling of the GGL measurement in upcoming data sets should take into account all these different effects, as it has been done already with recent data sets \citep{Prat_2022}. 

In this work, we will be concerned with a new paradigm of galaxy-galaxy lensing, one in which lens galaxy positions will correspond to galaxies that are at redshifts higher than source galaxies, and not lower (as it is the case in the traditional GGL configuration). As current and future imaging surveys acquire deeper photometry, we will be able to define galaxy samples at higher and higher redshifts \citep{HSC_dropouts_I_luminosity, HSC_dropouts_II_clustering, 2021arXiv210315862M, highz}, but measuring the positions of these galaxies is much easier than also measuring their shapes, and therefore many of these high redshift samples will only serve as \textit{lens} (position) samples in galaxy-galaxy lensing. In this way, more and more often we will have lens (position) galaxy samples at redshifts higher than source (shape) galaxy samples. To this new configuration of galaxy-galaxy lensing, where lens galaxies are behind source galaxies, we give the name inverse galaxy-galaxy lensing (IGGL).

This paper will explore this new IGGL measurement, its unique characteristics and differences with the traditional paradigm of GGL, and what information can we extract from it. In Section \ref{sec:theory}, we will describe the modeling of the galaxy-galaxy lensing measurement, point out the differences between GGL and IGGL, and present the simulated measurements that we will use for testing the different scenarios. In Section \ref{sec:dependencies}, we will explore the different sensitivities of GGL and IGGL with respect to model parameters, and the efficiency of the IGGL measurement as a function of lens and source redshift. In Section \ref{sec:model-constraints}, we will look into parameter constraints, and compare or combine them with other probes such as cosmic shear and galaxy clustering. Section \ref{sec:discussion} will present a discussion and Section \ref{sec:conclusions} will summarize our findings and present conclusions.

\section{Theory and simulated data}
\label{sec:theory}

\begin{figure*}
\begin{center}
\includegraphics[width=0.83\textwidth]{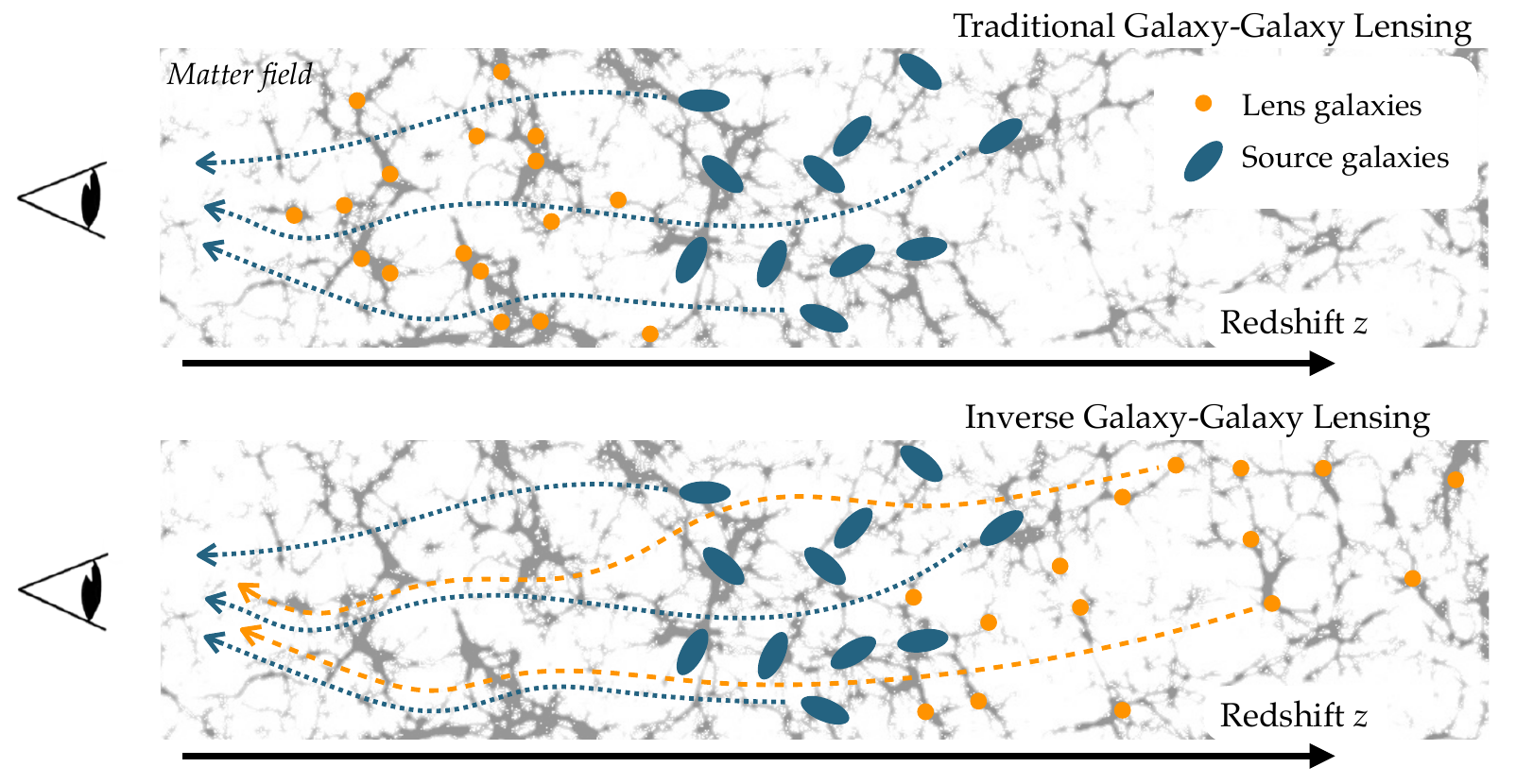}
\caption{\textit{Top panel}: Illustration of lensing in the traditional GGL setup, where the source (shape) sample is being lensed by the matter around the lens (position) sample. In this case, the GGL signal is dominated by the correlation between intrinsic lens galaxy density and source shear, the first term in Eq.~(\ref{eq:all_terms}). \textit{Bottom panel}: Illustration of the IGGL setup, with source galaxies being the same as in the upper panel, but lens (position) galaxies being at higher redshifts. In this case, the intrinsic lens galaxy density does not correlate with source shear, and therefore the first term in Eq.~(\ref{eq:all_terms}) vanishes. Instead, the lens sample density is being magnified by the same matter that shears the source galaxies, creating a magnification-shear correlation, and there will also be correlations due to intrinsic alignments. }
\label{fig:draw}
\end{center}
\end{figure*} 

\begin{figure}
\begin{center}
\includegraphics[width=0.5\textwidth]{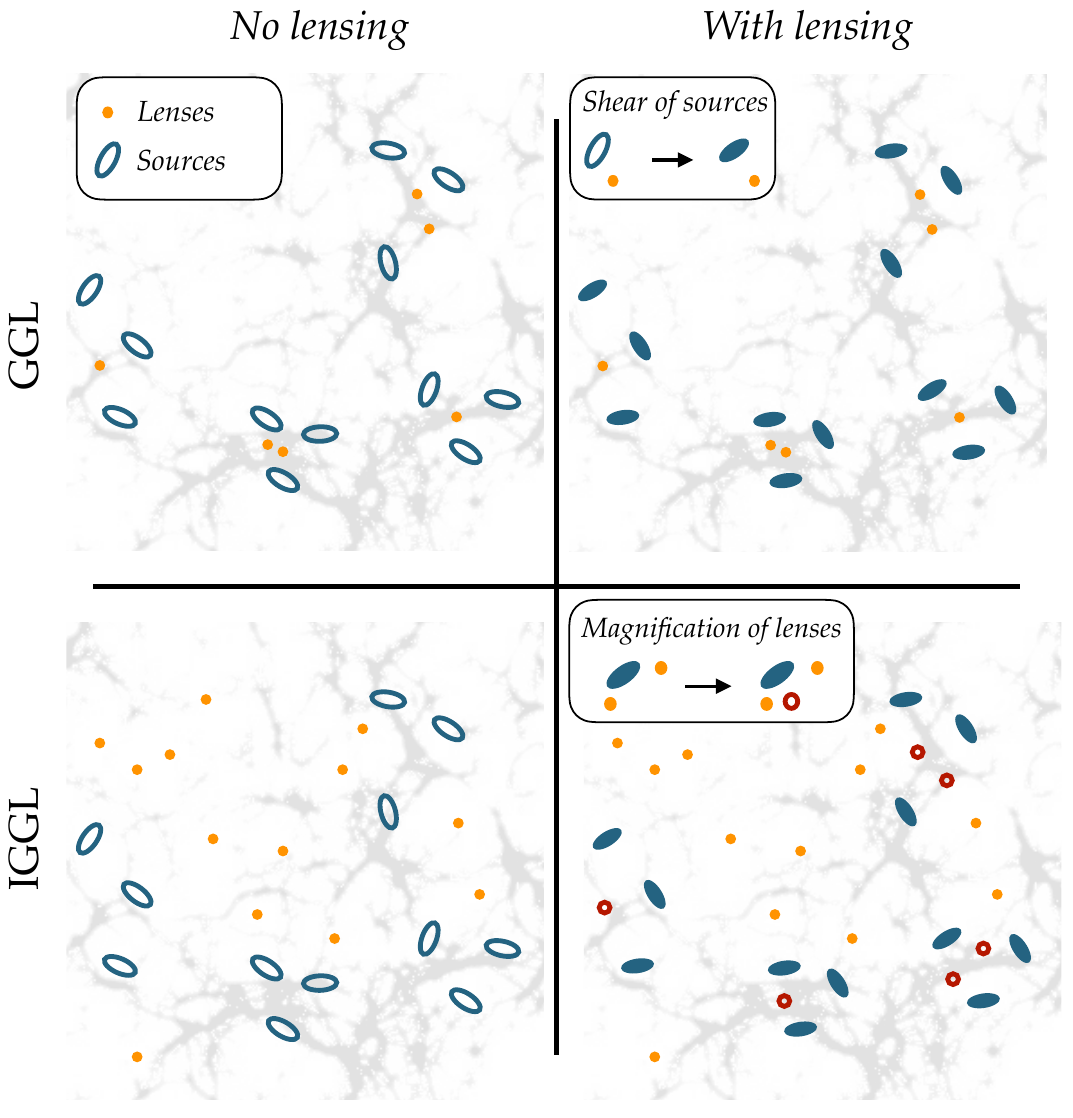}
\caption{Line-of-sight illustration of GGL (first row) and IGGL (second row). For GGL, the leading order effect on the image with and without lensing is the tangential shear on the shape of source galaxies (blue ellipses) due to the mass field (in grey). The mass field is traced by the lens (position) sample (orange dots), as they overlap in redshift, and therefore there is a correlation between intrinsic lens density and source shear. For IGGL, the shapes of source galaxies will be altered in the same way (as they are the same sample and the same mass distribution is causing the shear). The intrinsic density of lenses is not correlated with the underlying mass, but the lens (position) sample will be magnified, and the changes in density due to magnification will be correlated with mass and therefore with source shear. The correlation of these two effects on the lens and source samples represents the magnification term in Eq.~(\ref{eq:all_terms}). For this illustration, we have assumed $\alpha^\mathrm{mag} > 1$ and therefore a positive magnification-shear correlation.   }
\label{fig:draw2}
\end{center}
\end{figure} 

In this section we will describe the theory behind the galaxy-galaxy lensing observables used in this work, which will closely follow the modeling from \citep{Prat_2022}, together with the different simulated scenarios we will consider to explore the inverse galaxy-galaxy lensing case. 

\subsection{Modeling the tangential shear observable}
\label{sec:model}

To model the galaxy-galaxy lensing observable, we need to consider the cross-correlation between the observed lens galaxy density in the sky, $\delta_g^\text{obs}$, and the observed ellipticity of source galaxies, $e^{\text{obs}}$. For the former, we have combined effects of the intrinsic galaxy density and the change in the galaxy density due to weak lensing magnification from mass between the lens galaxy and the observer:
\begin{equation}
\label{eq:delta_g}
    \delta_g^\text{obs} =  \delta_g^\text{int} + \delta_g^\text{mag} .
\end{equation}
The change in lens galaxy density produced by magnification is proportional to the convergence \citep*{y3-2x2ptmagnification}, and therefore we have:
\begin{equation}\label{eq:definition_C}
\delta_g^\text{mag} = C^\mathrm{mag} \kappa_l \, , 
\end{equation}
where $\kappa_l$ is the convergence field at the lens redshift and $C^\mathrm{mag}$ is a proportionality factor, which we can separate into two terms, an area effect and a flux effect on the number density change: $C^\mathrm{mag} = C^\mathrm{mag}_\mathrm{area} + C^\mathrm{mag}_\mathrm{flux}$.  It can be shown that $C^\mathrm{mag}_\mathrm{area}= - 2$ \citep*{y3-2x2ptmagnification} while $C^\mathrm{mag}_\mathrm{flux}$ will depend specifically on the sample, and can be calculated in different ways using data, simulations and artificial galaxy injections \citep*{y3-2x2ptmagnification}. Given these two contributions, we will write $C^\mathrm{mag} = 2 (\alpha^\mathrm{mag} -1)$, where $\alpha^\mathrm{mag}$ is a property of the sample and is equivalent to $C^\mathrm{mag}_\mathrm{flux}/2$. In this way, Eq.~(\ref{eq:delta_g}) becomes: 
\begin{equation}
\label{eq:delta_g2}
    \delta_g^\text{obs} =  \delta_g^\text{int} + 2 (\alpha^\mathrm{mag} -1) \kappa_l.
\end{equation}
For source galaxies, we model their observed ellipticities including the contribution from shear $\gamma$ and intrinsic ellipticity \citep{Prat_2022}: 
\begin{align}
\label{eq:ellipticities}
e^{\text{obs}} = \gamma + e^{\text{int}},
\end{align}
where we are working in the weak lensing regime, $\gamma\ll 1$. Correlating $\delta_g^\text{obs}$ with $e^{\text{obs}}$ to get the relevant cross-correlation for GGL, we get four terms representing different physical effects:
\begin{equation}
\label{eq:all_terms}
\begin{split}
 &\left< \delta_g^\text{obs} e^{\text{obs}} \right> = 
 \left< (\delta_g^\text{int} +  2 (\alpha^\mathrm{mag} -1) \kappa_l) \,  (\gamma + e^{\rm int}) \right> = \\
 &= \underbrace{\left< \delta_g^\text{int} \gamma\right>}_{\text{trad GGL}} 
 + \, \underbrace{C^\mathrm{mag}\left< \kappa_l \, \gamma \right>}_{\text{lens mag}} + 
 \, \underbrace{\left< \delta_g^\text{int} e^{\rm int}\right>}_{\text{IA}} + 
 \, \underbrace{C^\mathrm{mag}\left< \kappa_l \, e^{\rm int} \right>}_{\text{lens mag + IA}}
   \end{split}
\end{equation}

It is now useful to discuss these four terms, as they will have different relative importance in GGL and IGGL. The first term of Eq.~(\ref{eq:all_terms}) represents the leading term in the traditional GGL measurement. As illustrated in the top rows of Figs.~\ref{fig:draw} and \ref{fig:draw2}, intrinsic lens galaxy density ($\delta_g^\text{int}$) traces the foreground mass distribution, which causes an average shear ($\gamma$) on the shape of background source galaxies. At intermediate and large scales, the relationship between the lens galaxy density and the mass distribution is modeled using a simple linear bias model, $\delta_g^\text{int} = b \delta_m$, and therefore the first term of Eq.~(\ref{eq:all_terms}) can be written as $\left< \delta_g^\text{int} \gamma\right> = b \left< \delta_m \gamma\right>$. The cosmological information in traditional GGL is contained in the term $\left< \delta_m \gamma\right>$, and therefore there is a strong degeneracy between galaxy bias and cosmological information, particularly with the amplitude of the matter power spectrum ($\sigma_8$). This is the main reason why GGL needs to be combined with galaxy clustering to probe cosmology. In the IGGL case, however, lens (position) galaxies are behind sources (at higher redshifts), and therefore the intrinsic lens galaxy density is not correlated with the shear of source galaxies, thus this term vanishes (see the graphical representation in the lower row of Figure \ref{fig:draw}). In this way, the leading term in GGL will vanish in IGGL, which will make the measurement independent of galaxy bias and amplify the effects of the previously second-order terms in traditional GGL. 


The second term in Eq.~(\ref{eq:all_terms}) represents the cross-correlation between the change in lens galaxy density due to magnification and the shear in the shape of source galaxies. This term is subdominant in traditional GGL \citep{Prat_2022}, but it will be very important for IGGL (given that, in general, we will have $\alpha^\mathrm{mag} \neq 1$). Figure \ref{fig:draw2} shows a graphical representation of the effect of lens magnification in IGGL. In the upper row, we see the effect of weak lensing in traditional GGL, where lens galaxies are tracing the matter field between source galaxies and us, and that matter field is producing a shear in the shapes of background source galaxies. In the lower row, for the case of IGGL, lens galaxies are not correlated with the matter field between source galaxies and us, since they are behind source galaxies (at higher redshifts). The effect of shear on source galaxies is the same, and now weak lensing also magnifies the light from distant lens galaxies, which changes the number density of galaxies in our sample. If $\alpha^\mathrm{mag} > 1$, positive changes in number density will occur in areas of higher matter density along the line of sight, which will correlate positively with tangential shear (this is the case for the example of Fig.~\ref{fig:draw2}). If $\alpha^\mathrm{mag} < 1$, a negative correlation will be observed. 

The third term of Eq.~(\ref{eq:all_terms}) corresponds to the cross-correlation between intrinsic lens galaxy density and intrinsic galaxy shapes, which is the main contribution of intrinsic alignments (IA) to the GGL signal. For the IA modeling, we will consider the NLA (Non-linear Linear Alignment \citep{Hirata2004,2007NJPh....9..444B}) and TATT (Tidal Alignment and Tidal Torquing, \citep{Blazek_2019}) models. It is important to note that this third term of Eq.~(\ref{eq:all_terms}) will only be non-zero if the lens and source populations overlap in redshift, since the lens/source pairs must be physically associated to create an IA signal. The last term in Eq.~(\ref{eq:all_terms}) corresponds to the cross-correlation between IA and lens magnification, and this can be non-zero even if the lens and source populations do not overlap in redshift. 

Given all four terms in Eq.~(\ref{eq:all_terms}), one can then write the total contribution to the GGL tangential shear observable as a function of angular separation in the sky $\theta$, $\gamma_t(\theta)$. Following \citep{Prat_2022}, we obtain, for the curved sky projection ($\alpha = \alpha^\mathrm{mag}$ for brevity): 

\begin{align}
\begin{split}
\gamma_t(\theta) &= \sum_\ell \frac{2\ell+1}{4\pi\ell(\ell+1)}P^2_\ell(\cos\theta) \notag \\  
&\times \left[ C_{gm}(\ell) + 2 (\alpha -1) \left(C_{mm}(\ell)+ C_{mI}(\ell)\right) + C_{gI}(\ell) \right],
\end{split}
\end{align}
where the different $C_{ij}(\ell)$ are the angular cross-power spectra of the different tracers: lens galaxies $g$, matter $m$ and intrisic alignments $I$ (see \citep{Prat_2022,2022PhRvD.105h3529S} for more details on these terms). As a difference with respect to \citep{Prat_2022}, we compute the matter power spectrum including the effect of baryonic feedback, using the prescription of \citep{HMCODE}, and we will use the $T_\mathrm{AGN}$ parameter to vary the strength of the effect. The full model will be specified by $\Lambda$CDM cosmological parameters, galaxy bias, baryonic feedback, magnification and IA parameters, and nuisance parameters concerning mutliplicative shear bias and redshift calibration for both lens and source populations. Finally, for the case of traditional GGL, as in \citep{Prat_2022}, we will use the point-mass marginalization technique to remove small-scale information (mixing non-linear galaxy bias and baryonic effects) when we run Monte Carlo Markov Chains (MCMCs) using our GGL simulated data, following the work in \citep{MacCrann_2019}. We will give further details and specify the model choices made in each test throughout this paper. In order to implement these expressions, we use the \textsc{CosmoSIS} framework \citep{Zuntz_2015} to compute all the theoretical calculations. 

\subsection{Simulated data}
\label{sec:sims}

The theoretical modeling described above can be used to generate simulated data, given a complete set of model parameters and redshift distributions for the lens and source galaxy populations. For this paper, where we will explore the differences and dependencies of GGL and IGGL, we want to produce simulated measurements that are characteristic of these cases, and for that we will assume different redshift configurations and parameter values. In particular, we will explore three different scenarios: 
\begin{itemize}
    \item Traditional GGL: This will be a standard case of GGL, where lens galaxies are at lower redshift than source galaxies. 
    \item Inverse GGL (ideal): For this ideal IGGL case, we will use the same source galaxy sample as in the case above but we will use a narrow lens sample at higher redshift, without overlap in redshift between lenses and sources. 
    \item Inverse GGL (realistic): This IGGL case will have a broader lens redshift distribution at lower redshifts, which will create redshift overlap between lenses and sources. This will be a more realistic example of IGGL in real data.  
\end{itemize}

\begin{figure}
\begin{center}
\includegraphics[width=0.45\textwidth]{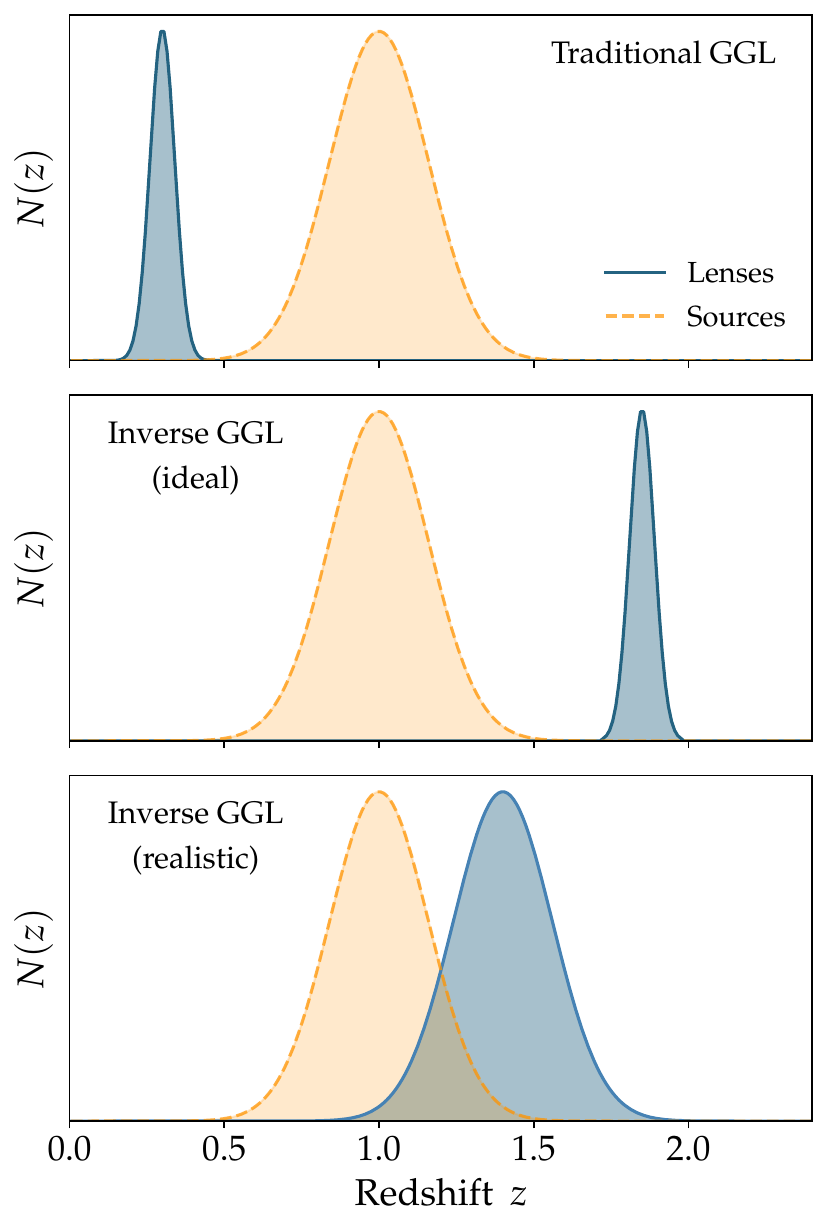}
\caption{Redshift distributions for the lens and source galaxy populations of the three different GGL configurations. The top panel shows the traditional GGL setup, where the lens (position) sample (blue) has a narrow redshift distribution at lower redshifts than the source (shape) sample (orange). The middle panel shows an ideal IGGL setup, where the lens sample is at a higher redshift than the source sample. This setup is ideal because of the narrow redshift distribution of the lens sample, which is less realistic as redshift distribution width tends to increase as a function of redshift. The bottom panel shows a more realistic setup, where the lens sample shows a wider redshift distribution, including redshift overlap between lenses and sources. }
\label{fig:nzs_theory}
\end{center}
\end{figure} 

The redshift distributions for these three different cases are depicted in the three different rows of Fig.~\ref{fig:nzs_theory}. The source redshift distributions are Gaussians $\mathcal{N}(\mu,\sigma) = \mathcal{N}(1,0.16)$, and the lens redshift distributions for the three cases are $\mathcal{N}(0.3,0.04)$, $\mathcal{N}(1.85,0.04)$ and $\mathcal{N}(1.4,0.16)$, respectively. The three simulated measurements are all generated at the same $\Lambda$CDM cosmology, assuming no IA, and with different values of galaxy bias ($b=1.4$ for GGL, $b=2.0$ for both IGGL cases) and magnification coefficients ($\alpha^\mathrm{mag} = 0.2$ for GGL, $\alpha^\mathrm{mag} = 2.0$ for both IGGL cases). We choose galaxy bias and magnification parameters for lens galaxies to grow with redshift as that is the trend we observe in the data \citep{Prat_2022,y3-2x2ptmagnification}. Even if these simulated measurements are made at fixed parameter values, in the next section we will explore the impact of varying model parameters. 

In order to get the covariance of the simulated measurements, we use the Gaussian covariance approximation \citep{2021MNRAS.508.3125F} with Rubin LSST Y1-like specifications: $f_\mathrm{sky} = 0.3$, source galaxy density $n^s_g = 2.5$ arcmin$^{-2}$, shape noise $\sigma_e = 0.3$ and lens galaxy densities of $n^l_g = 0.3$ arcmin$^{-2}$ for the traditional GGL and IGGL ideal cases (the narrow redshift distributions) and $n^l_g = 1.0$ arcmin$^{-2}$ for the broader lens sample in the realistic IGGL case. Given these simulated $\gamma_t(\theta)$ measurements and covariances, which are made for $\theta \in (1,500)$ arcmins in 27 angular bins, we can estimate the signal-to-noise of each of the three cases. Using the entire range of angular scales, we get signal-to-noise estimates of S/N = 299, 36 and 60 for traditional GGL, ideal IGGL and realistic IGGL, respectively. When using only angular scales over 8$h^{-1}$ Mpc at the lens redshift, the respective values change to S/N = 62, 19 and 31. The S/N is computed here and elsewhere in the paper as $\sqrt{\chi^2 - N_{\text{dp}}}$, where $\chi^2= \gamma_{t} \mathbf{C}^{-1} \gamma_{t}$, $N_{\text{dp}}$ is the number of data points in the measurement and $\mathbf{C}^{-1}$ is the inverse covariance. The simulated measurements for the three cases, together with their error bars, are shown as grey point with error bars in Fig.~\ref{fig:param_exploration}, which will be discussed in detail in the next section. 

\section{Exploring IGGL: Features and Dependencies}
\label{sec:dependencies}

In this part we will use the theory model and the simulated data discussed in the previous section to explore the differences between traditional GGL and inverse GGL, in terms of model parameter dependencies and lens-source geometry.  

\subsection{Parameter dependencies}
\subsubsection{Magnification, IA, cosmology and galaxy bias}

\begin{figure*}
\begin{center}
\includegraphics[width=0.98\textwidth]{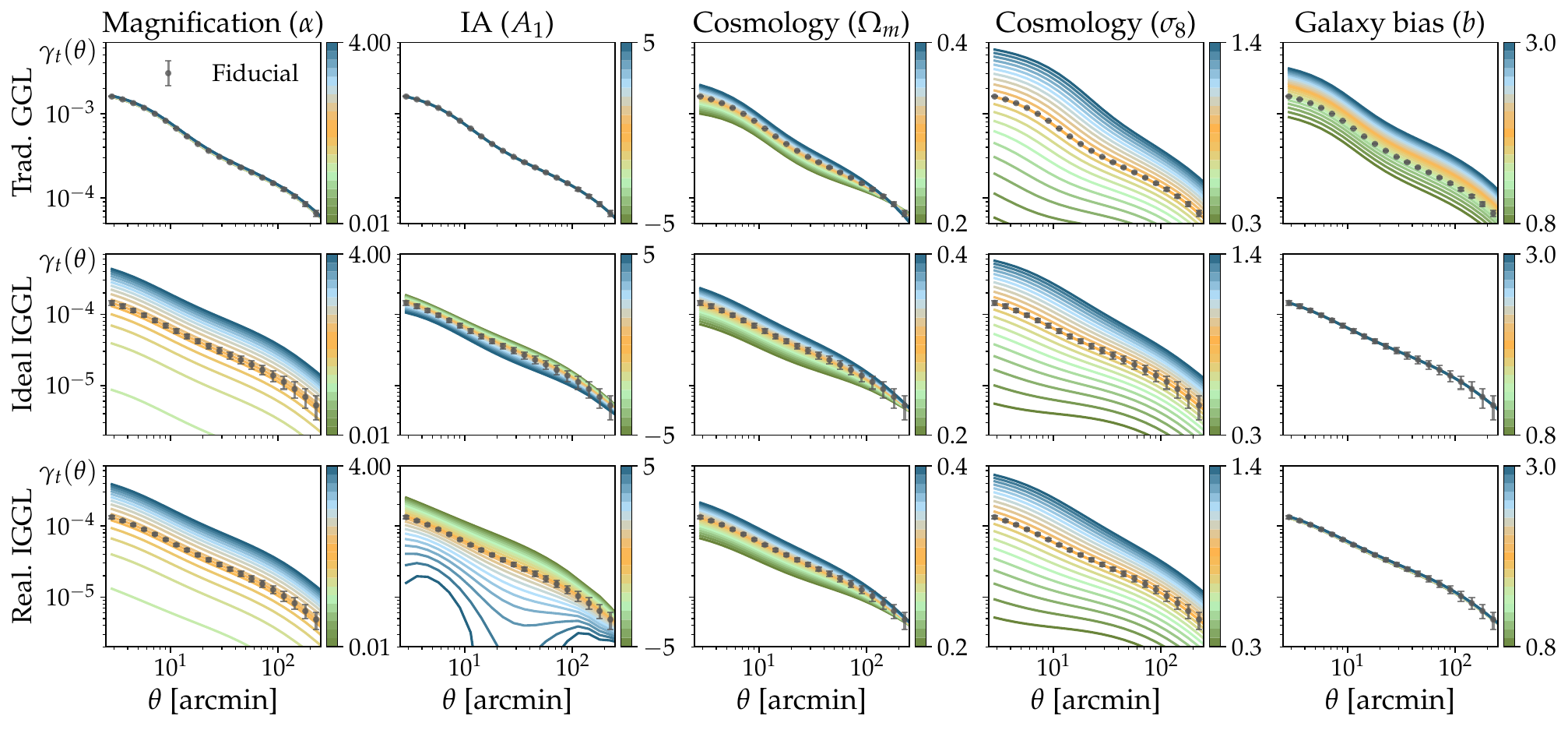}
\caption{Exploration of the effect of different model parameters in traditional and inverse (both ideal and realistic) GGL. The redshift distributions considered for the three cases (three rows) are those depicted in Fig.~\ref{fig:nzs_theory}. The fiducial simulated measurements and errors, described in \S \ref{sec:sims}, are the dark dots with corresponding error bars. These are compared to the model variations regarding the magnification coefficient, the amplitude of the tidal alignment in the IA model, cosmological parameters and galaxy bias, represented as lines color-coded by the values of the parameter being ranged over. As expected, the IGGL cases are much more sensitive to magnification and IA than the traditional GGL case. IGGL is not significantly affected by galaxy bias, unlike GGL, but is still sensitive to cosmology in a similar manner.    }
\label{fig:param_exploration}
\end{center}
\end{figure*} 

\begin{figure*}
\begin{center}
\includegraphics[width=0.98\textwidth]{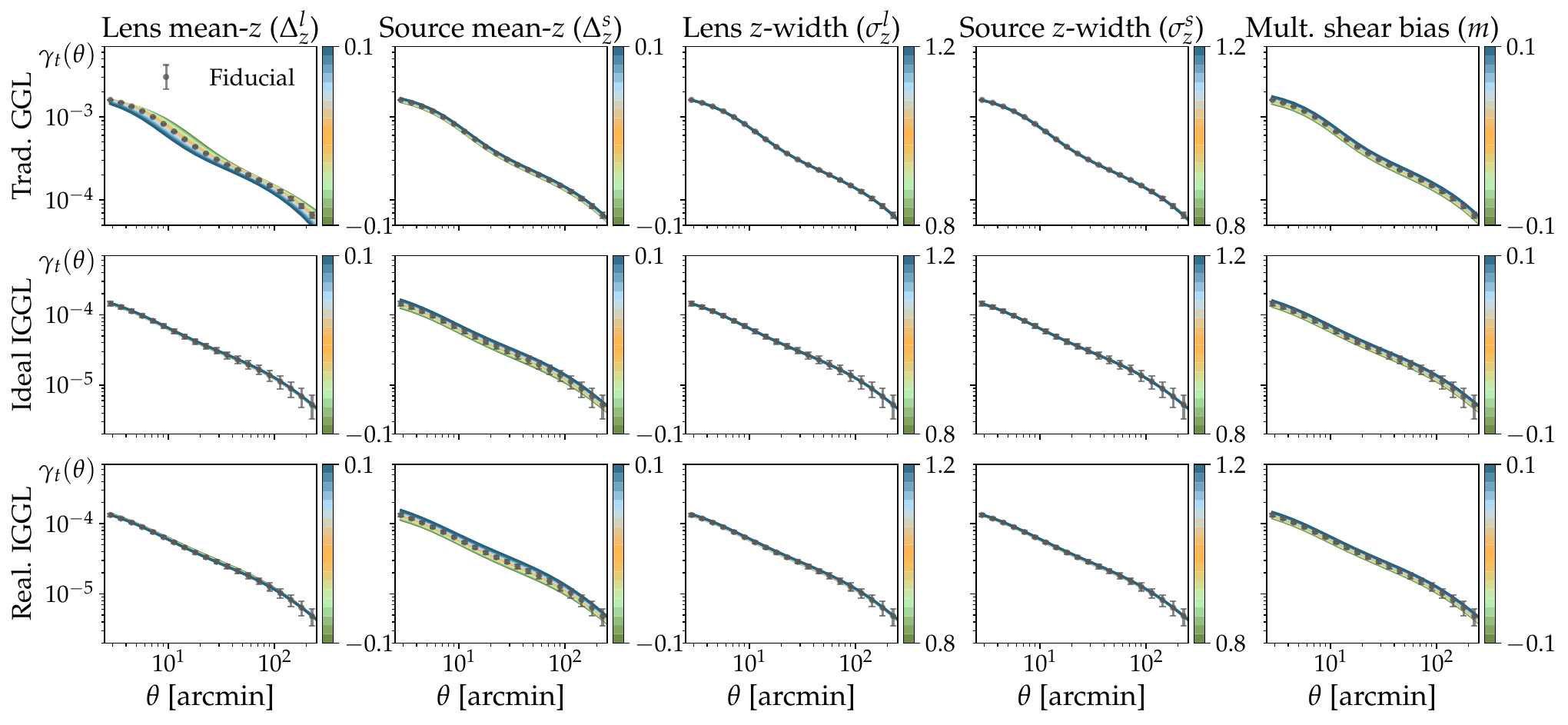}
\caption{Analogous plot to that of Fig.~\ref{fig:param_exploration}, but now showing the variation of nuisance parameters regarding redshift distributions of lens and source galaxies (means and widths) and multiplicative shear bias. The range of the nuisance parameters is chosen to be wider than it will be constrained by priors in parameter inference, and even so the dependence is shown to be much smaller than that of magnification, IA and cosmological parameters in Fig.~\ref{fig:param_exploration}. }
\label{fig:param_exploration2}
\end{center}
\end{figure*}

Figure \ref{fig:param_exploration} shows the simulated measurements described in \S\ref{sec:sims} as grey points with error bars, and then explores the dependence of those simulated measurements on several of the model parameters described also in \S\ref{sec:model}. The three rows correspond to the three cases we are considering, traditional GGL, ideal IGGL and realistic IGGL, with redshift distributions depicted in Fig.~\ref{fig:nzs_theory}. The different columns show the variation of different model parameters: magnification ($\alpha^\mathrm{mag}$, or simply $\alpha$), IA (the amplitude of tidal alignment in the NLA and TATT models, $A_1$), cosmological parameters ($\Omega_m$, $\sigma_8$), and galaxy bias ($b$).  

The first column in Fig.~\ref{fig:param_exploration} shows the impact of varying the lens magnification parameter $\alpha$, as defined in Eq.~(\ref{eq:all_terms}). Compared to its negligible impact on the traditional GGL data vector (first row), altering $\alpha$ has a much more significant impact on the data vector in both IGGL cases (middle and lower row). This is to be expected for a few reasons. First, the traditional GGL case has lower-redshift lens (position) galaxy samples, thus the lensing efficiency of magnification is much lower than for the IGGL case. Second, magnification is a second-order effect in the traditional GGL data vector, whereas in the IGGL case, it is leading order, as discussed in section \S\ref{sec:model}. A graphical representation of the lens magnification contribution to IGGL is shown in Fig.~\ref{fig:draw2}. 

The second column in Fig.~\ref{fig:param_exploration} shows the impact of varying the effect of IA, parametrized here by the amplitude of the tidal alignment in the NLA model \citep{Hirata2004,2007NJPh....9..444B}, $A_1$, which is the parameter that will affect the GGL observable the most. Similar to the case of magnification, we see how IA has negligible impact on the traditional GGL case (first row), here because there is no overlap in redshift between lenses and sources in that case. For IGGL, we see a mild impact of IA for the ideal case (middle row) and a much stronger one for the realistic case (lower row). This is because the ideal IGGL case has no redshift overlap between lenses and sources, but it gets IA dependence through the cross-correlation between IA and magnification, represented by the last term in Eq.~(\ref{eq:all_terms}). The realistic IGGL case does have redshift overlap between lenses and sources, and therefore it gets contributions from the pure IA term in Eq.~(\ref{eq:all_terms}), in addition to the cross-correlation with magnification. 

The third and fourth column of Fig.~\ref{fig:param_exploration} show the impact of varying cosmological parameters $\Omega_m$ and $\sigma_8$ in the three cases considered, and the dependence is comparable for all cases (the dependence is stronger with the amplitude of matter fluctuations, $\sigma_8$). 

A key difference between GGL and IGGL is shown in the last column of Fig.~\ref{fig:param_exploration}, where we show the impact of varying the galaxy bias parameter for lens galaxies. Traditional GGL shows a strong dependence, similar to that of the $\sigma_8$ parameter, illustrating the reason why cosmological information is degenerate with galaxy bias in traditional GGL. On the other hand, both IGGL cases show a negligible impact of galaxy bias variations, which will be crucial in allowing IGGL to constrain cosmology and other astrophysical parameters free of any degeneracies with galaxy bias. 


\subsubsection{Redshift and shear calibration}

Figure \ref{fig:param_exploration2} follows the same structure as Fig.~\ref{fig:param_exploration}, but in this case showing the dependence of the three observables on nuisance parameters related to redshift and shear calibration. In particular, we are showing the dependence on the mean and width of the lens and source redshift distributions, as well as the impact of multiplicative shear bias. From the first column, we can see how IGGL is less sensitive to lens redshifts than traditional GGL, while the other columns show similar dependencies. Regardless, the range of the parameters is chosen to be ample, wider than it will be constrained by priors in parameter inference (which we will study in \S \ref{sec:model-constraints}), and even so the dependence is shown to be much smaller than that of magnification, IA and cosmological parameters in Fig.~\ref{fig:param_exploration}.  


\subsubsection{Baryonic effects}

\begin{figure}
\begin{center}
\includegraphics[width=0.45\textwidth]{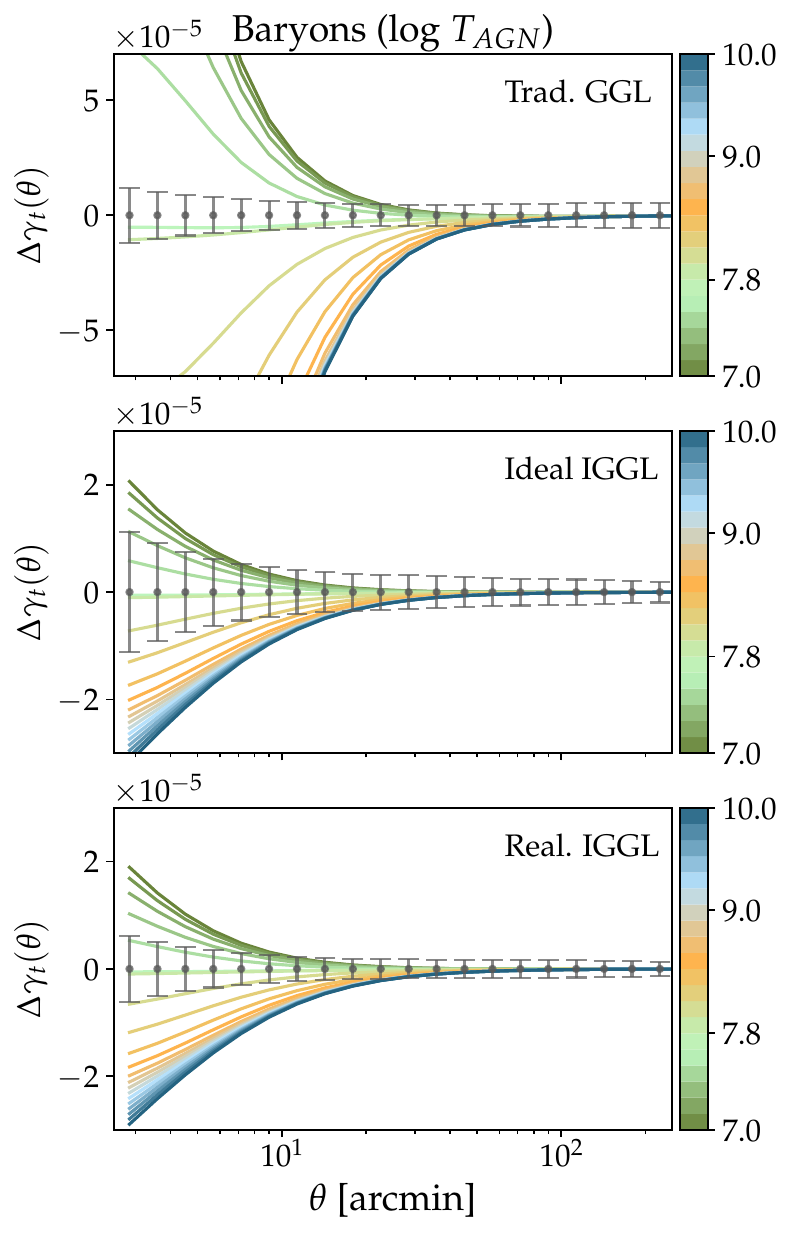}
\caption{Exploration of the effect of baryons on GGL and IGGL. In all three panels, the measurement errors of 
the data vector for GGL, ideal IGGL, and realistic IGGL, as described in \S \ref{sec:sims}, are illustrated with gray error bars. The lines represent the difference between the fiducial data vector and the new data vector with the value of log $T_{\mathrm{AGN}}$ changed (marked by the color bar). In traditional GGL, baryonic effects, parameterized by log $T_{\mathrm{AGN}}$ in the HMcode-2020 model \citep{HMCODE}, are highly degenerate with non-linear galaxy bias at small scales. Because IGGL has no galaxy bias dependence, smaller scales can be used to place constraints on $T_{\mathrm{AGN}}$ free of any degeneracy with non-linear galaxy bias. As seen above, both the ideal and realistic IGGL cases (as introduced in Fig.~\ref{fig:nzs_theory}) do still vary significantly with different values of $T_{\mathrm{AGN}}$.  }
\label{fig:baryons}
\end{center}
\end{figure}

While the linear power spectrum is analytically mod-
eled to high precision \citep{CAMB,CLASS}, calculations of the non-linear power spectrum at smaller scales require complex,
simulation-based models. Corrections to the non-linear power spectrum in current surveys include two effects: baryonic feedback and galaxy bias. Earlier in this Section, we discussed how Fig.~\ref{fig:param_exploration} showed
that IGGL had negligible dependence on galaxy bias.
This fact has important implications in that we can explore the usage of small angular scales in IGGL without
worrying about non-linear galaxy bias. However, if we
go to small scales, we will need to consider the impact
of baryonic effects on the matter power spectrum, and
therefore on our observables.

Baryonic effects on dark matter halo structure are modeled by feedback from high-energy processes such as supernovae and active galactic nuclei (AGNs), alongside other processes such as the evolution of non-collisional star systems. These non-gravitational phenomena are theorized to alter dark matter halo structure, specifically by causing disturbances in their cores, consequently altering the matter-power spectrum at small scales (see, e.g.~\citep{OWLs1,BAHAMAS}). In past cosmological surveys, there have been multiple parameterizations of baryonic feedback, such as halo concentration, star formation, supernova feedback, and chemical enrichment to name a few (see, e.g., \citep{mead2015, Harnois2015,HALOFIT,HMCODE}). 

While some models of baryonic feedback are preferred over others, it is not clear which model reflects the data because the baryonic feedback signal is quite degenerate with galaxy bias at small scales. Because of the complicated nature of these effects and their degeneracies with galaxy bias, some analyses have simply thrown out data at scales smaller than a certain threshold \citep{throwout1, planck2015, DESy1}, while others have used methods to correct the power spectrum at relevant scales \citep{y3csbaryons, DSbaryons, KiDSbaryons}). 

To determine just how much of an effect baryonic feedback has on the IGGL data vector, we use the HMCode-2020 model \citep{HMCODE} to model baryonic effects in the matter power spectrum at small scales. In this model, all the effects of baryonic feedback on the power spectrum are combined into one single `AGN temperature' parameter, log $T_\mathrm{AGN}$, which governs the strength of the feedback. Figure \ref{fig:baryons} shows the dependence of the observables we consider on that parameter. For all three observables, the effect of baryonic feedback is only relevant at small scales, and increasing log $T_\mathrm{AGN}$ results in a suppression of the matter power spectrum and therefore a reduced $\gamma_t$ signal. This is expected, as the redistribution of baryonic mass in halos through processes like AGN feedback and star formation leads to a suppression of the matter power spectrum on small scales. From the figure, we can see how the impact of the effect is larger for traditional GGL, but it also has a significant effect on IGGL at small scales. While GGL is more affected by the $T_\mathrm{AGN}$ parameter, its signal will be degenerate with the galaxy bias parameter, and thus we expect that the measurement of the parameter will have lower constraining power than the IGGL case.

\subsubsection{Negative $\gamma_t$ values}

\begin{figure}
\begin{center}
\includegraphics[width=0.48\textwidth]{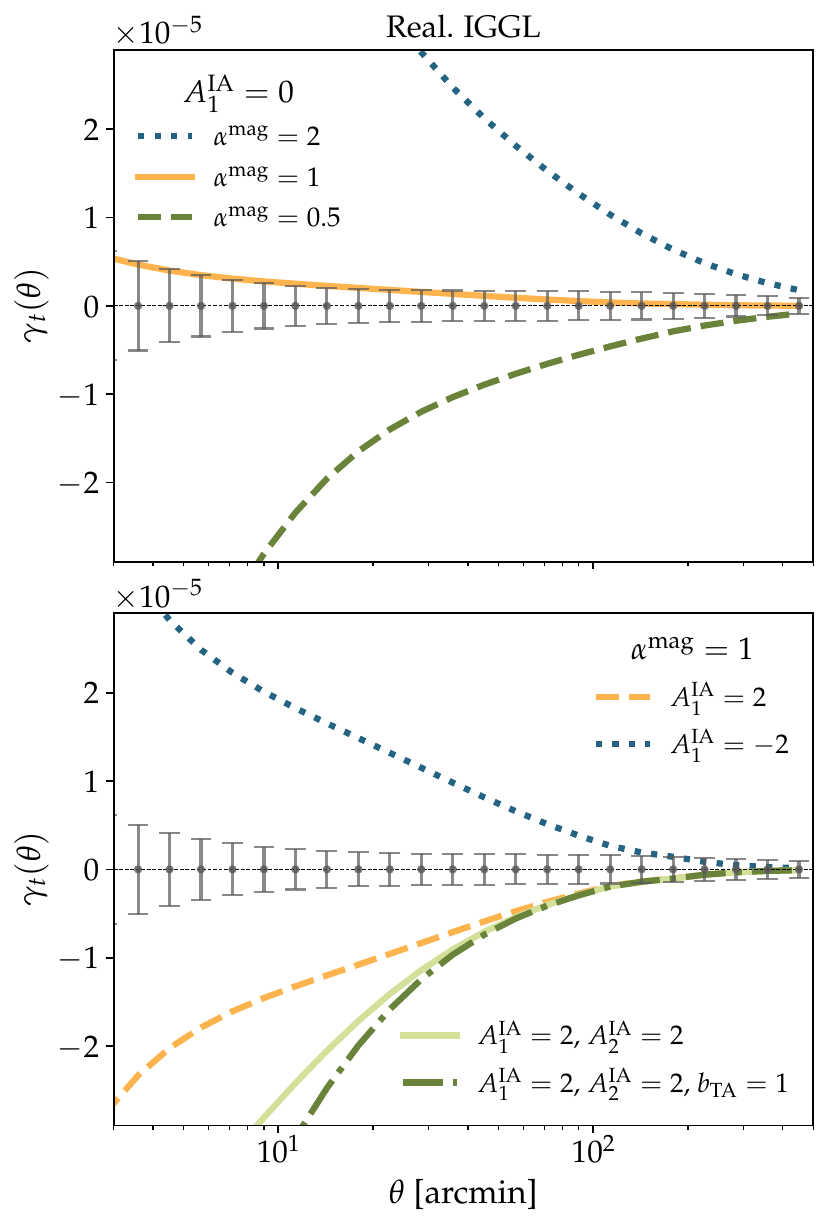}
\caption{Exploration of the conditions that yield a negative IGGL data vector (only considering the realistic IGGL case). In the top panel, it is shown that an $\alpha^\mathrm{mag}$ parameter of less than 1 can give a negative data vector, as expected from Eq.~(\ref{eq:all_terms}), and that $\alpha^\mathrm{mag} = 0.5$ yields a negative signal at high signifcance, in the absence of IA. In the bottom panel, in the absence of magnification ($\alpha^\mathrm{mag} = 1$), it is shown that positive values of $A_1^\mathrm{IA}$ and $A_2^\mathrm{IA}$ in the TATT model can give a negative data vector as well, and so do positive values for the bias of the tidal alignment component ($b_\mathrm{TA}$). Because the traditional GGL term in Eq.~(\ref{eq:all_terms}) is always positive around galaxies and leading order, IGGL is the first GGL scenario where a negative data vector is possible when measured around galaxies. }
\label{fig:negative}
\end{center}
\end{figure} 

One other feature that would distinguish IGGL from traditional GGL is the possibility of observing negative $\gamma_t (\theta)$ values in the measurement. In traditional GGL, the signal is expected to be positive when measured around galaxies, and negative correlations are only observed when measuring lensing around cosmic voids or similar tracers of matter underdensities \citep{Krause_2012,Melchior_2014,2013ApJ...762L..20K,2017MNRAS.465..746S,2016MNRAS.455.3367G}. However, IGGL can present negative correlations in different realistic scenarios. Looking at Eq.~(\ref{eq:all_terms}), we can see how $\alpha^\mathrm{mag} < 1$ will make the magnification contributions negative, and also a positive $A_1$ in IA can produce negative GGL measurements. 

Figure \ref{fig:negative} shows a few different cases of negative $\gamma_t$ signals in the realistic IGGL scenario (with the redshift distributions depicted in the lower panel of Fig.~\ref{fig:nzs_theory}). The errorbars of the gray points centered at zero are there to indicate the size of the measurement uncertainties. The upper panel shows three different $\gamma_t$ signals for three different values of the magnification parameter $\alpha^\mathrm{mag}$, in the case of no IA. For $\alpha^\mathrm{mag} = 2$, we get a positive $\gamma_t$ signal for IGGL. For $\alpha^\mathrm{mag} = 1$, the contribution of magnification vanishes, and in this case of no IA the remaining signal corresponds to the small impact of traditional GGL (due to the overlap of the lens and source redshift distributions in the lower panel of Fig.~\ref{fig:nzs_theory}). For $\alpha^\mathrm{mag} = 0.5$, there is a significant negative $\gamma_t$ signal for IGGL. 

The lower panel of Fig.~\ref{fig:negative} shows different IGGL $\gamma_t$ signals, for different parameter values of the TATT (Tidal Alignment and Tidal Torquing, \citep{Blazek_2019}) IA model, in the case of no lens magnification. In GGL, a positive value of the amplitude of tidal alignment amplitude $A_1^\mathrm{IA}$ yields a negative $\gamma_t$ signal. Similarly, positive values for the tidal torquing $A_2^\mathrm{IA}$ component make the signal even more negative, and so do positive values for the bias of the tidal alignment component ($b_\mathrm{TA}$). On the other hand, negative values of the amplitude of tidal alignment amplitude $A_1^\mathrm{IA}$ yield a positive $\gamma_t$ signal.

\subsection{Dependence on lens-source geometry and efficiency}
\label{sec:geometry}

\begin{figure}
\begin{center}
\includegraphics[width=0.5\textwidth]{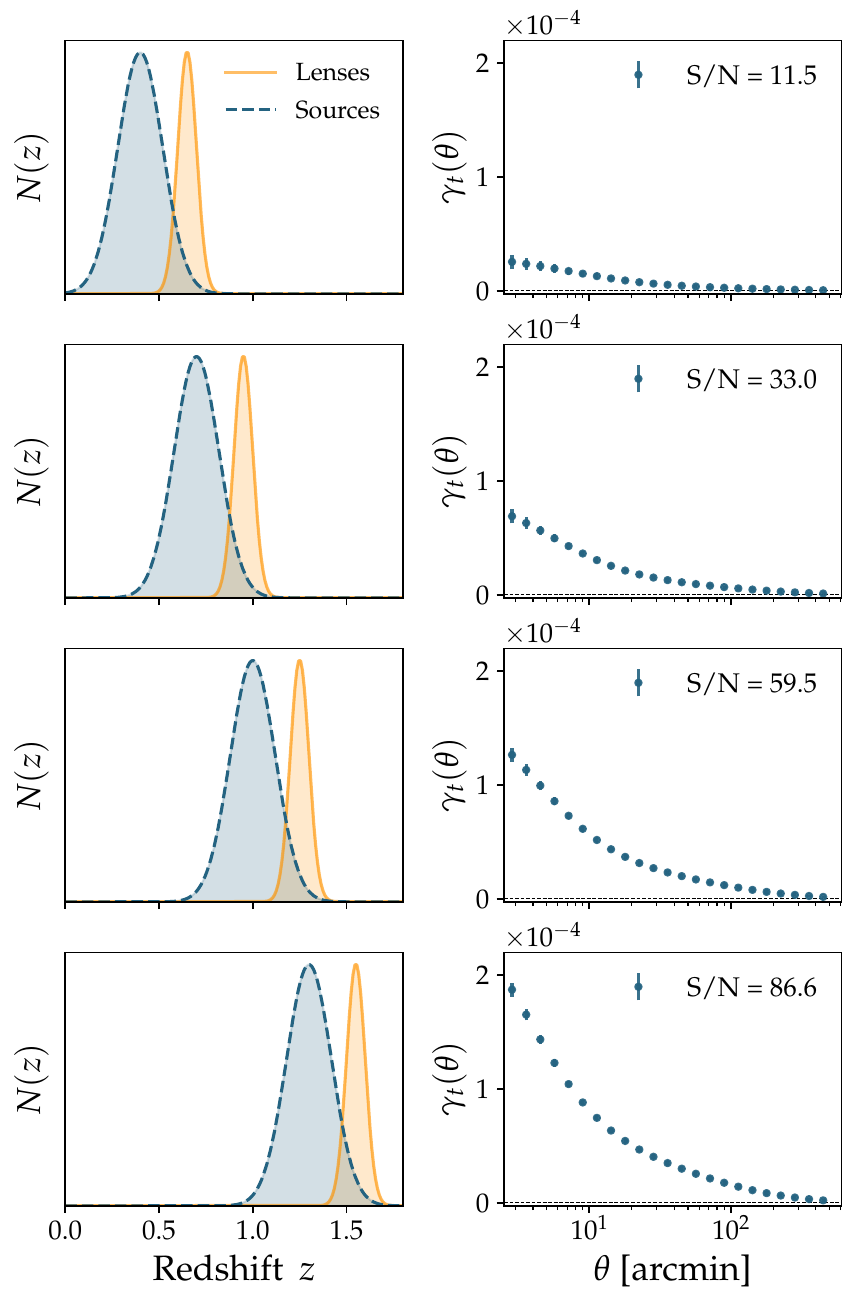}
\caption{Change in measurement signal-to-noise for different IGGL redshift configurations, as described in \S \ref{sec:geometry} (S/N is computed in every case for the entire angular range, from 1 to 500 arcmins). As the lens (position) and source (shape) samples go to higher redshifts, the signal becomes stronger, as there is more mass between the observer and the lens/source samples, which increases the strength of the shear and magnification signals (the two multiplying lensing kernels get higher with higher redshifts). }
\label{fig:geometry1}
\end{center}
\end{figure} 

\begin{figure*}
\begin{center}
\includegraphics[width=0.9\textwidth]{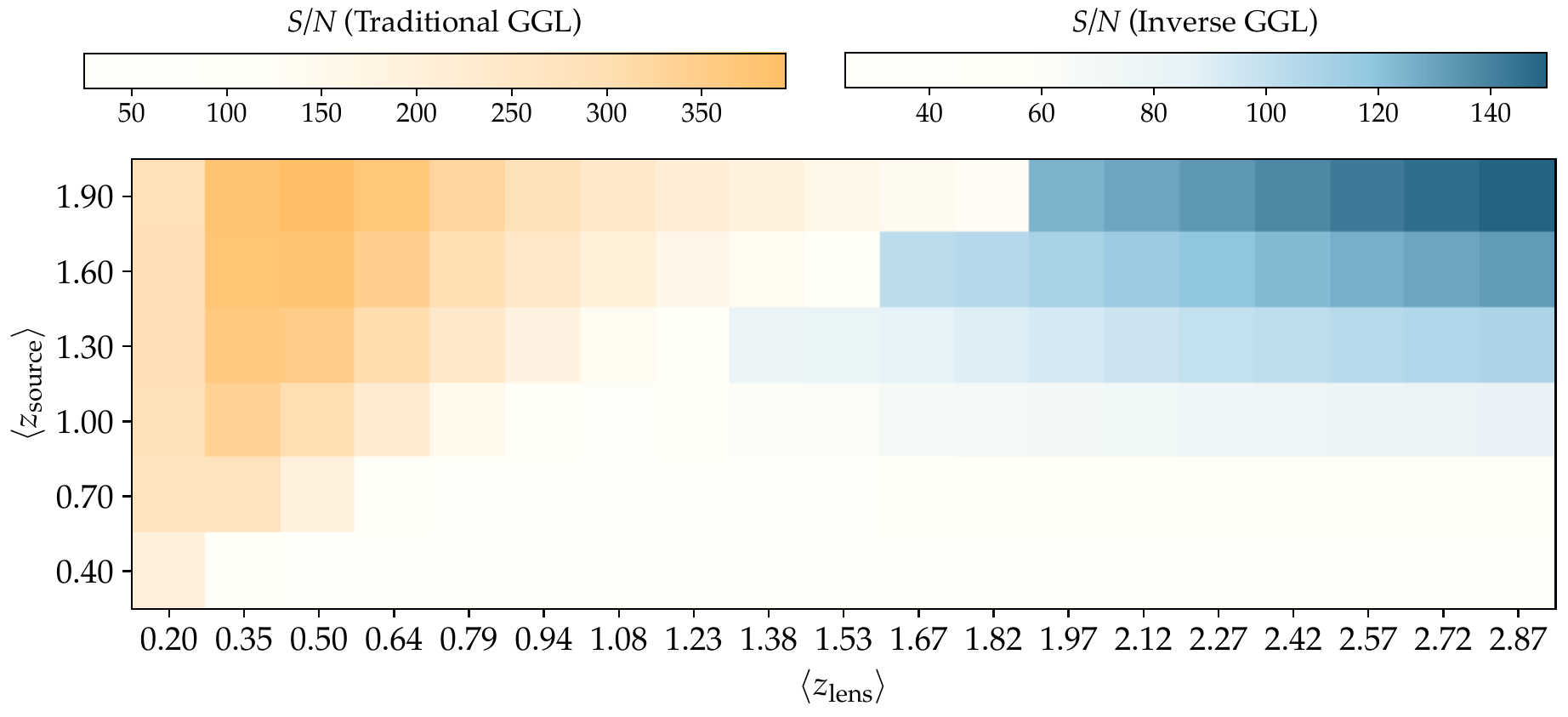}
\caption{Two-dimensional analysis of the change in GGL signal-to-noise for different redshift configurations. We consider lens and source redshift distributions with the same width, densities, and model parameters as those in Fig.~\ref{fig:geometry1}, but now centered at different combinations of lens and source redshifts, represented by their horizontal and vertical positions, respectively.  Colors show the S/N of $\gamma_t$ for every one of those cases. The traditional GGL regime can be seen in the upper left corner of the figure ($z_\mathrm{lens} < z_\mathrm{source}$), and the IGGL regime ($z_\mathrm{lens} > z_\mathrm{source}$) is found in the right side of the plot. We have used a different color bar scale for GGL and IGGL (with the same lower S/N limit), as we expect a higher S/N reach for traditional GGL. The region around $z_\mathrm{lens} = z_\mathrm{source}$ where the $\gamma_t$ S/N is at a minimum, serves as a way to separate the GGL and IGGL regimes. Also, while traditional GGL S/N peaks when lenses are about halfway between the observer and sources, IGGL S/N increases with lens and source redshifts.}
\label{fig:geometry2}
\end{center}
\end{figure*} 

So far we have explored the ideal and realistic IGGL cases with the redshift distributions considered in Fig.\ref{fig:nzs_theory}. In both of those cases, the source redshift distribution is at a mean redshift of $z=1$, and the \emph{lens} redshift distribution is at higher redshifts. In this part we want to explore the efficiency of the IGGL measurement as a function of lens (position) and source (shape) redshifts. 

As a first test, Fig.~\ref{fig:geometry1} shows simulated IGGL measurements for different lens and source $N(z)$ configurations, fixing the relative redshift position between them but moving them together higher in redshift as we go down the rows of the plot. In order to make this test only about geometry, for every row in the plot the galaxy density for lens and source galaxy populations is fixed ($n^l_g = 1.0$ arcmin$^{-2}$, $n^s_g = 2.5$ arcmin$^{-2}$), and so are the parameters related to lens magnification ($\alpha^\mathrm{mag} = 2.0$) and IA ($A_1^\mathrm{IA} = A_2^\mathrm{IA} = 0$). In the figure, we can observe how the S/N of $\gamma_t$ increases significantly as both redshift distributions are shifted to higher redshifts. This is because for IGGL to be efficient through lens magnification we need both lens and source distributions to be at high redshifts, so that there is enough matter between those distributions and us to be able to cause shear distortions on the sources and lens magnification on the lenses. Also, note that this is a conservative case, because
in reality the galaxy density and potentially the magnification coefficients would increase with redshift, making the S/N redshift trend even stronger.

As a second test, to probe the more general dependence of IGGL with lens and source redshifts, we consider lens and source redshift distributions with the same width, densities, and model parameters as those in Fig.~\ref{fig:geometry1}, but now centered at different combinations of source and lens redshifts. Figure \ref{fig:geometry2} shows the S/N of $\gamma_t$ for every one of those cases, from which we can draw several conclusions. First of all, since the figure shows the $\gamma_t$ S/N systematically for varying lens and source redshifts, it includes the traditional GGL case in the upper left corner of the figure ($z_\mathrm{lens} < z_\mathrm{source}$). Then, the IGGL case ($z_\mathrm{lens} > z_\mathrm{source}$) is found in the right side of the plot, and we have used a different color bar scale for GGL and IGGL (with the same lower S/N limit), as we expect a higher S/N reach for traditional GGL. It is important to note how there is a region around $z_\mathrm{lens} = z_\mathrm{source}$ where the $\gamma_t$ S/N is at a minimum, and this serves as a natural way to separate the GGL and IGGL scenarios in a clear way. The plot also shows that while traditional GGL shows a peak of S/N around $0.35 < z_\mathrm{lens} < 0.5$, where the lens redshift distribution overlaps the most with the source lensing kernel, for IGGL there is no such S/N peak: IGGL S/N will always increase with lens and source redshifts, because the S/N will be given by the product of two lensing kernels, and those will always grow with redshift. Finally, we note that the plot in Fig.~\ref{fig:geometry2} was made with no IA in the models, but adding realistic values of the IA parameters does not change the overall conclusions from this plot.

\section{Model constraints from IGGL and comparison to other probes} \label{sec:model-constraints}

In this part we are interested in placing model constraints given the $\gamma_t (\theta)$ two-point functions of GGL described in \S\ref{sec:theory}, in order to analyze which parameters or parameter combinations they are sensitive to. Later in this section, for the purposes of comparison to IGGL, we will also be interested in model constraints from other two-point functions such as cosmic shear and galaxy clustering, and therefore we will now describe our inference scheme for general two-point functions. Overall, given our model $M$, we want to infer parameters $\mathbf p$ from the set of measured two-point correlation functions, $\hat{\mathbf D}$.
The theoretical model prediction for the two-point correlation functions, computed using the parameters $\mathbf p$ of the model ${M}$, is $\mathbf T_M(\mathbf{p})$. We compare the measurements and model predictions using a Gaussian likelihood, using the data covariance, $\mathbf{C}$, defined above:
\begin{equation}
\mathcal{L}(\hat{\mathbf D}|\mathbf p, M) \propto e^{-\frac{1}{2}\left[\left(\hat{\mathbf D}-\mathbf T_M(\mathbf p)\right)^{\mathrm{T}} \mathbf{C}^{-1}\left(\hat{\mathbf D}-\mathbf T_M(\mathbf p)\right)\right]}.
\end{equation}
In this way, the posterior probability distribution for the parameters $\mathbf p$  of the model $M$ given the data $\hat{\mathbf D}$ is given by
\begin{equation}
P(\mathbf p |\hat{\mathbf D}, M) \propto \mathcal{L}(\hat{\mathbf D}|\mathbf{p}, {M})P(\mathbf{p}|{M}),
\end{equation}
where $P(\mathbf{p}|{M})$ is the prior probability distribution on the parameters. The specific priors and the allowed ranges on every model parameter will be discussed specifically in each of the cases. 

\subsection{Model constraints from IGGL and comparison to traditional GGL}
\label{sec:IGGL_constraints}

\begin{table}
\caption{The model parameters and their priors used in the GGL and IGGL MCMC chains, and some of the other chains in \S \ref{sec:model-constraints} (differences are discussed in their respective sections). The parameters are defined in \S \ref{sec:IGGL_constraints}.}
\begin{center}
\renewcommand{\arraystretch}{1.3}
\begin{tabular*}{\columnwidth}{ l  @{\extracolsep{\fill}} c  c}
\hline
\hline
Parameter & \multicolumn{2}{c}{Prior}  \\  
\hline 
\multicolumn{2}{l}{{\bf Cosmology}} \\
$\Omega_{\mathrm{m}}$  &  Flat  & (0.1, 0.9)  \\ 
$10^{9}A_{\mathrm{s}}$ &  Flat  & ($0.5,5.0$)  \\ 
$n_{\mathrm{s}}$ &  Flat  & (0.87, 1.07)  \\
$\Omega_{\mathrm{b}}$ &  Flat  & (0.03, 0.07)  \\
$h$  &  Flat  & (0.55, 0.91)   \\
$m_{\nu}$ & Flat & (0.06, 0.6) \\
\hline
\multicolumn{2}{l}{{\bf Galaxy Bias} } \\
$b$   & Flat  & (0.8, 4.0) \\
\hline
\multicolumn{2}{l}{{\bf Lens magnification} } \\
$\alpha^\mathrm{mag}$ (GGL) & Gaussian &  ($0.20, 0.05$) \\
$\alpha^\mathrm{mag}$ (IGGL) & Gaussian &  ($2.0, 0.1$) \\
\hline
\multicolumn{2}{l}{{\bf Baryonic feedback} } \\
log $T_\mathrm{AGN}$ (GGL) & Fixed &  (7.8) \\
log $T_\mathrm{AGN}$ (IGGL) & Flat &  (7.0, 10.0) \\
\hline
\multicolumn{2}{l}{{\bf Lens redshifts} } \\
$\Delta z^l$ (GGL) & Gaussian &  (0, 0.0065) \\
$\Delta z^l$ (IGGL) & Gaussian &  (0.0, 0.012) \\
\hline
\multicolumn{2}{l}{{\bf Source redshits} } \\
$\Delta z^s$ & Gaussian &  (0.0, 0.004) \\
\hline
\multicolumn{2}{l}{{\bf Intrinsic alignments} } \\
$A_1^\mathrm{IA}$, $\alpha_1^\mathrm{IA}$  & Flat &  (-5.0, 5.0) \\
$A_2^\mathrm{IA}$, $\alpha_2^\mathrm{IA}$, $b_\mathrm{TA}$  & Fixed &  (0.0) \\
\hline
\multicolumn{2}{l}{{\bf Shear calibration} } \\
$m$ & Gaussian &  (0.0, 0.013) \\
\hline
\hline
\end{tabular*}
\end{center}
\label{tab:params}
\end{table}

Now we look at model constraints from the IGGL realistic case that we have been considering in the paper, presented in \S\ref{sec:theory}, and we will compare these constraints to those coming from the traditional GGL case, described in the same section. For traditional GGL, we utilize only angular scales over 8 $h^{-1}$Mpc at the lens redshift, and we employ the point-mass marginalization technique to remove smaller-scale information at the time of running our MCMCs, following the work in \citep{MacCrann_2019}. We follow this approach in traditional GGL because both non-linear galaxy bias and baryonic effects are present at small scales \citep{Prat_2022}. Because we cut out small scales and use point-mass marginalization for our traditional GGL MCMCs, we fix the value of $T_\mathrm{AGN}$ to the true value used to generate our test data vector, as there will be no sensitivity to it. For IGGL, however, there is no dependence on non-linear galaxy bias, as we have seen, and therefore we use smaller scales (over 1 $h^{-1}$Mpc) but we allow the baryonic model to vary in the MCMC chains for IGGL. Therefore, the IGGL case uses more angular scales but also varies one more parameter (log $T_\mathrm{AGN}$) in the model than the GGL case. 

Next we list the different parameters that are varied in our model, and their associated ranges and priors (which are also summarized in Table \ref{tab:params}): 

\begin{itemize}
    \item \textbf{Cosmological parameters}: Throughout this paper, we will consider a flat $\Lambda$CDM cosmological model. The six cosmological parameters we vary are listed in Table \ref{tab:params}, together with their respective uniform priors. These prior ranges are chosen to encompass at least five times the 68\% C.L. from relevant external constraints (these are the same cosmological parameters and priors considered in the DES Y3 Fiducial analysis \citep{y3-3x2ptkp}). Also, even though we sample the amplitude of primordial scalar density perturbations $A_{\mathrm{s}}$, we will often show results with the amplitude of density perturbations at $z=0$ in terms of the RMS amplitude of mass on scales of $8h^{-1}$ Mpc in linear theory, $\sigma_8$. In addition to these cosmological parameters, we will also vary parameters related to galaxy bias, lens magnification, IA, baryonic feedback, redshift and shear calibration. 

    \item \textbf{Redshift and shear calibration}: The choice of priors related to redshift and shear calibration is inspired by the LSST Year 1 case in \cite{Zhang_2022}, where there is a prescription for those priors as a function of redshift. In particular, the uncertainty in the mean redshift of the lens galaxy distribution follows the form $\sigma(\Delta z^l) = 0.005(1+z)$, the uncertainty in the mean redshift of the source galaxy distribution follows the form $\sigma(\Delta z^s) = 0.002(1+z)$, and the uncertainty on shear multiplicative bias parameter is $\sigma(m) = 0.013$. This establishes our choice of redshift and shear calibration priors, given the mean redshifts of our lens and source galaxy distributions. 

    \item \textbf{Lens magnification parameter}: $\alpha^\mathrm{mag}$ describes the sign and amplitude of the lens magnification effect, as in Eq.~(\ref{eq:delta_g2}). The choice of values for $\alpha^\mathrm{mag}$ is described in \S\ref{sec:sims}, for the lens galaxies considered in GGL and IGGL. In this section, we use Gaussian priors for them, so that we have $\alpha^\mathrm{mag} \sim \mathcal{N}(0.20,0.05)$ for GGL (lenses at $z=0.3$) and $\alpha^\mathrm{mag} \sim \mathcal{N}(2.0,0.1)$ for realistic IGGL (lenses at $z=1.4$). The values chosen for $\alpha^\mathrm{mag}$ are motivated by previous data analyses \cite{y3-2x2ptmagnification,highz}. 
    
    \item \textbf{(Linear) Galaxy bias parameter}: $b$ represents the relation between the underlying dark matter density field and the intrinsic galaxy density field, as defined in \S\ref{sec:theory}. For every case, we let $b$ vary with a uniform prior between 0.8 and 4.0.  

    \item \textbf{Baryonic feedback}: We use HMCode-2020 \citep{HMCODE} to model baryonic effects in the matter power spectrum, parametrized by the log $T_\mathrm{AGN}$ parameter. For the case of IGGL, where we go to smaller angular scales, we let log $T_\mathrm{AGN}$ vary with a uniform prior in the range (7.0, 10.0). For traditional GGL, where we remove small-scale information, the baryonic feedback parameter remains fixed at its true value, log $T_\mathrm{AGN} = 7.8$. 
    
    \item \textbf{Intrinsic Alignment (IA) parameters}: Our fiducial IA model is the NLA model, which is a subset of the TATT model. TATT has 5 parameters: two amplitudes governing the strength of the alignment for the tidal and for the torque part, respectively, $A_1^\mathrm{IA}$, $A_2^\mathrm{IA}$, two parameters modeling the dependence of each of the amplitudes in redshift, $\alpha_1^\mathrm{IA}$, $\alpha_2^\mathrm{IA}$, and $b_{\mathrm{TA}}$, describing the tidal alignment bias. As we said, we will use the NLA IA model, which corresponds to TATT with $A_2^\mathrm{IA} = 0$, $\alpha_2^\mathrm{IA} = 0$, and $b_{\mathrm{TA}}=0$. In this case, we only vary the parameters concerning tidal torquing, with broad uniform priors in the range (-5.0, 5.0). 

\end{itemize}

\begin{figure}
\begin{center}
\includegraphics[width=0.5\textwidth]{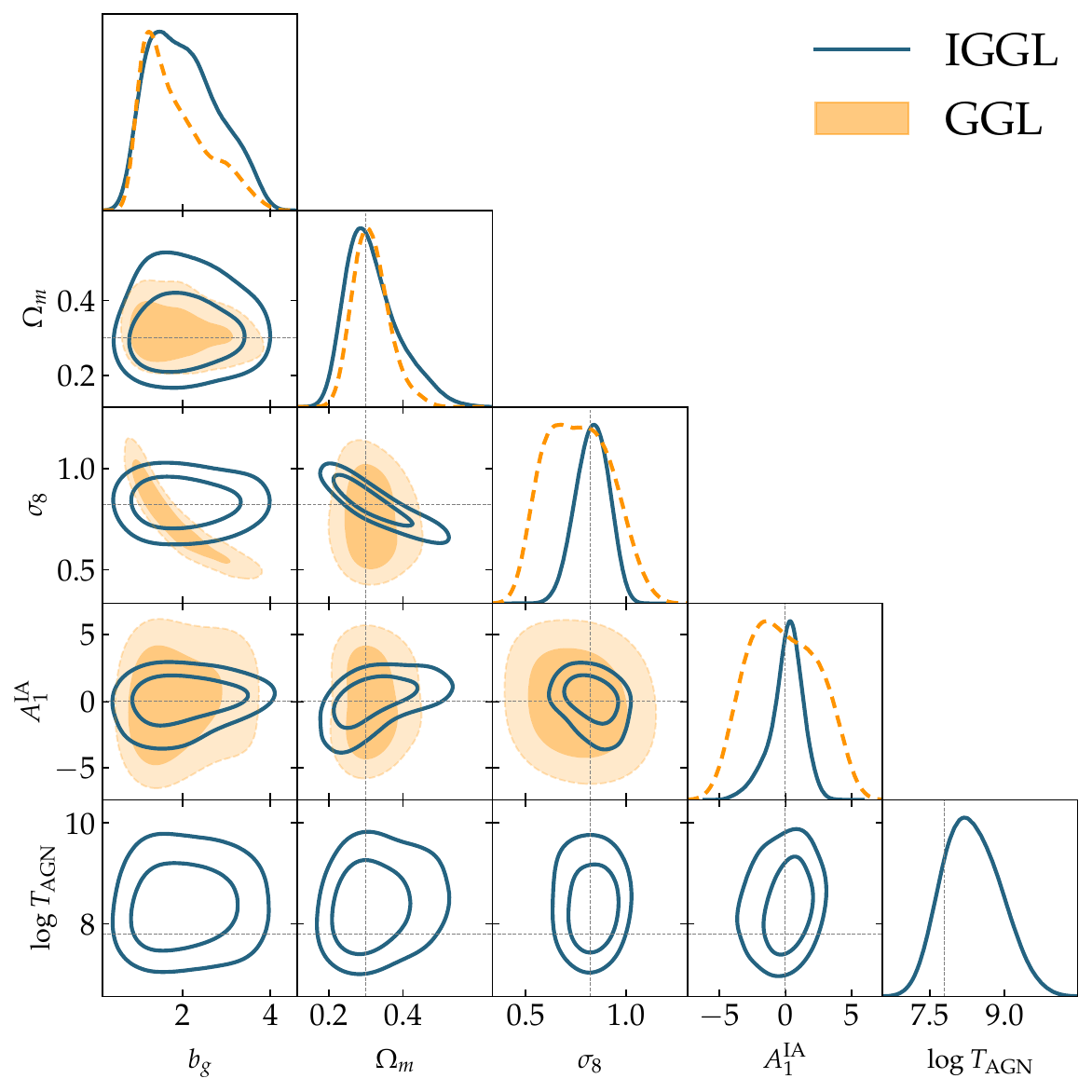}
\caption{Parameter constraints from GGL and realistic IGGL data vectors, as described in \S\ref{sec:IGGL_constraints}. The $\Omega_m$ parameter constraints are comparable, but IGGL provides significantly tighter constraints on $\sigma_8$, which is not degenerate with galaxy bias in IGGL (it is for tradiational GGL, as expeected). IGGL also obtains tighter constraints on the amplitude of tidal alignment in the IA model. Because IGGL is independent of galaxy bias, we use smaller scales and free the baryonic feedback parameter, $T_\mathrm{AGN}$, also obtaining a constraint on it.  }
\label{fig:ggl_iggl}
\end{center}
\end{figure} 

Figure \ref{fig:ggl_iggl} shows the model constraints from traditional GGL and realistic IGGL for a selection of the most relevant model parameters: galaxy bias $b$, the matter density $\Omega_m$, the amplitude of matter fluctuations $\sigma_8$, the IA amplitude in the NLA model $A_1^\mathrm{IA}$, and the baryonic feedback parameter log $T_\mathrm{AGN}$. Other varied parameters are not shown as they are either not constrained or dominated by the assumed priors, as described in Table \ref{tab:params}. From the figure, we can see how none of the measurements can place tight constraints on galaxy bias, but we see that traditional GGL shows a strong degeneracy between galaxy bias and $\sigma_8$. This is expected (see discussion in \S\ref{sec:model}), and it is the reason why cosmological inference using traditional GGL requires the combination with galaxy clustering in order to break this degeneracy \citep{y3-2x2ptaltlensresults}. On the other hand, IGGL does not depend on galaxy bias and therefore does not show a degeneracy between $b$ and $\sigma_8$, but it shows a degeneracy between $\sigma_8$ and $\Omega_m$, sometimes summarized in the $S_8 \equiv \sigma_8 (\Omega_m /0.3)^{0.5}$ parameter, characteristic of the pure lensing measurements like cosmic shear \citep{y3-cosmicshear1, y3-cosmicshear2}. This is again expected: cosmic shear is a lensing-lensing correlation (shear-shear) and so is IGGL, or at least its leading contribution (magnification-shear). 

In terms of constraining power, while they both constrain $\Omega_m$ similarly, IGGL provides much stronger constraints on $\sigma_8$ compared to GGL. IGGL is again much more sensitive to $A_1^\mathrm{IA}$, since there is significant overlap in redshift between lenses and sources (see last row of Figure \ref{fig:nzs_theory}), in addition to the cross-term between IA and magnification. Furthermore, we are varying the baryonic feedback parameter for IGGL, so we also get a constraint on that parameter which we do not get in traditional GGL (where we do not use small-scale information and therefore log $T_\mathrm{AGN}$ remains fixed). 

There is one important parameter that we are not showing in this plot, and that is the parameter concerning lens magnification, $\alpha^\mathrm{mag}$. This parameter is not relevant for GGL, but it is important for IGGL (see Fig.~\ref{fig:param_exploration}). The reason we do not show it is because it is significantly informed by our choice of prior (see Table \ref{tab:params}). However, even with that choice of prior, IGGL does bring some new information: using the fiducial prior of $\alpha^\mathrm{mag} = 2.0\pm 0.1$ we obtain a posterior of $\alpha^\mathrm{mag} = 1.996\pm 0.082$. 

\subsubsection{Dependence on IA and magnification priors}

\begin{figure}
\begin{center}
\includegraphics[width=0.48\textwidth]{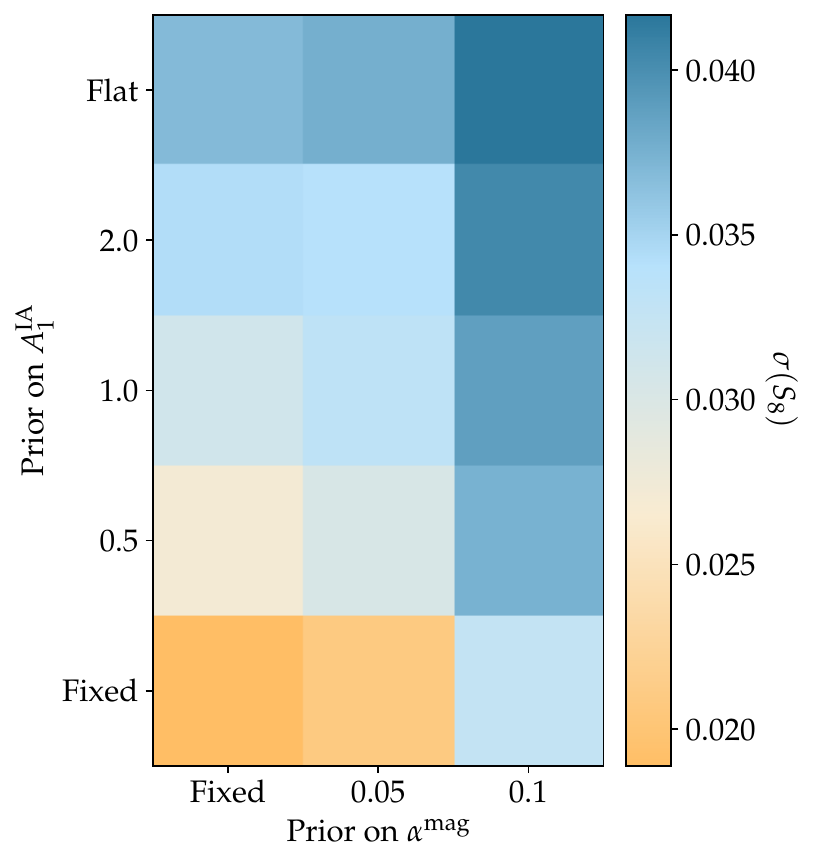}
\caption{Constraining power on $S_8$ from realistic IGGL, as a function of priors on magnification and IA. These 15 boxes represent 15 MCMCs where the priors on IA parameter $A_1$ and magnification parameter $\alpha^\mathrm{mag}$ have differing prior values, and the color represents the corresponding posterior uncertainty on $S_8$. For the values of the priors, 'Flat' represents a uniform, non-informative prior, 'Fixed' corresponds to the parameter not being varied, and the numbers are the $\sigma$ of Gaussian priors. From the figure, we see that we can improvements of more than a factor of two in $S_8$ constraining power if we tighten up the IA and magnification priors.}
\label{fig:S8_IA_mag}
\end{center}
\end{figure}

Fig. \ref{fig:ggl_iggl} shows the IGGL model constraints given our fiducial prior choices. Now we turn to exploring the limitations of IGGL in terms of prior choices, especially those related to magnification and IA. For magnification, the fiducial choice is a 5\% Gaussian prior, motivated by the findings in \citep{highz}, but it is possible that measurements of magnification coefficients using source injection techniques \citep{balrog} yield even stronger priors in the future. For IA, our fiducial choice of priors is a uniform, non-informative prior, but this situation can also be improved in the future with models like NLA and TATT being applied to multiple weak lensing datasets \citep{Blazek_2019, Blazek_2015, Blazek2012, blazek2011, Zhang_2010, DESy1, Hildebrandt2018, asgari_2021, y3-cosmicshear1,y3-cosmicshear2}. For these reasons, we believe our fiducial set of priors is realistic for LSST Y1, but it can potentially be more informative. Figure \ref{fig:S8_IA_mag} shows the IGGL cosmological constraining power, focusing on the $S_8 \equiv \sigma_8 (\Omega_m /0.3)^{0.5}$ parameter, given three choices of magnification priors and five choice of IA priors. From the figure, we see that we can get improvements of more than a factor of two in $S_8$ constraining power if we tighten up the IA and magnification priors.

\subsection{Comparison and combination with cosmic shear}
\label{sec:cosmic_shear}

In this section we will compare (and later combine) IGGL with cosmic shear, the angular two-point correlation between source galaxy shapes. To do this, we produce cosmic shear simulated data in analogy to the GGL simulated data described in \S\ref{sec:sims}. The cosmic shear measurement consists of the $\xi_+$  correlation function (see \citep{y3-cosmicshear1, y3-cosmicshear2} for a detailed description of the correlation function) with the same angular bins produced for the GGL measurements, using the same source bin and source number density. The S/N of the cosmic shear simulated measurement is S/N = 150, compared to S/N = 60 for IGGL. We use this cosmic shear simulated measurement to place constraints on the same model parameters described in Table \ref{tab:params}, using the same source priors (except for the parameters describing lens properties, which are not relevant for cosmic shear). The cosmic shear model constraints use the same minimum physical scale used for IGGL (1 $h^{-1}$Mpc), providing a fair comparison between the cosmic shear and IGGL constraints. 

\begin{figure*}
\begin{center}
\includegraphics[width=0.75\textwidth]{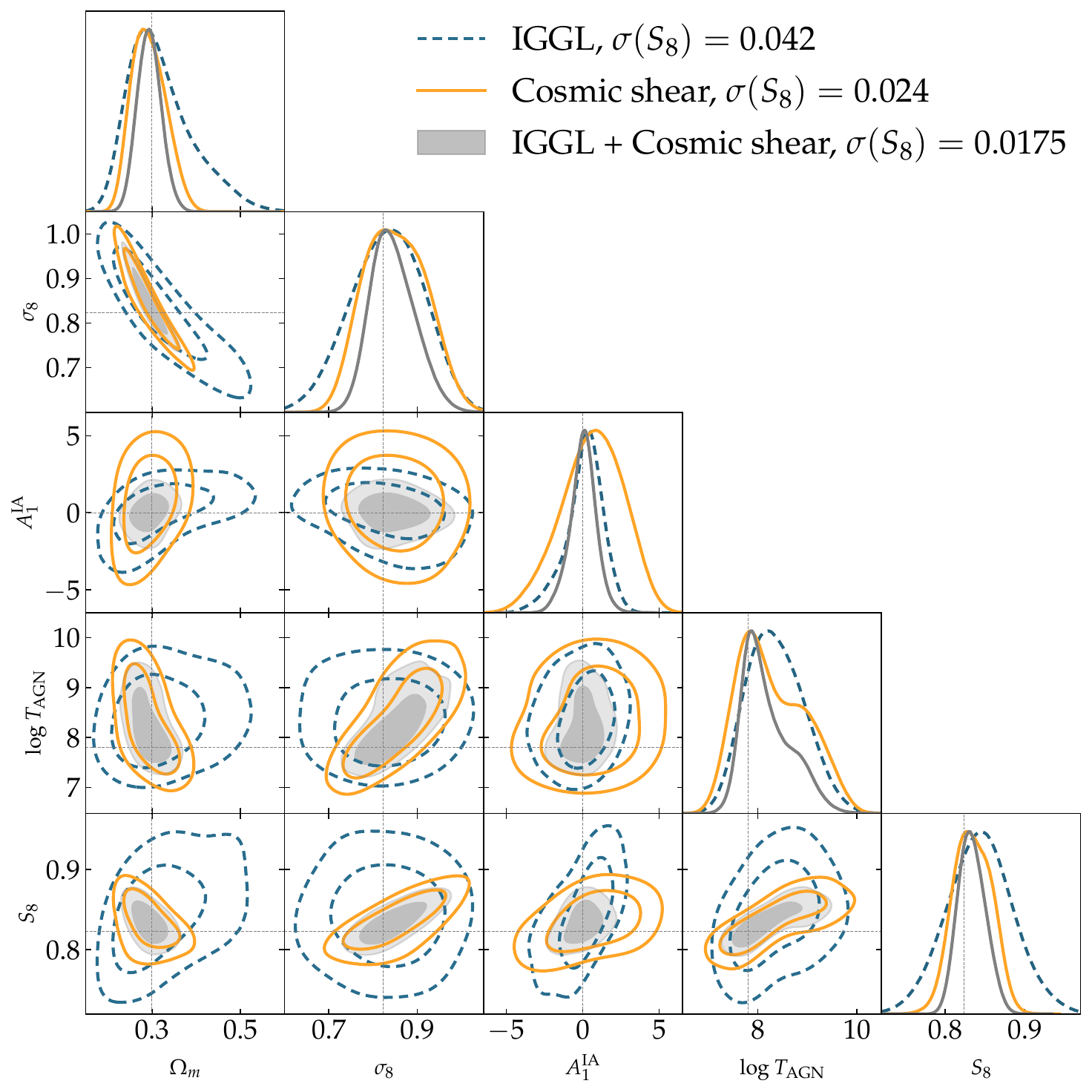}
\caption{Model constraints from (realistic) IGGL and cosmic shear measurements, and the combination of them, as described in \S \ref{sec:cosmic_shear}. As expected from signal-to-noise (S/N = 150 for cosmic shear, compared to S/N = 60 for IGGL) cosmic shear has  tighter constraints on cosmological parameters, but IGGL shows tighter constraints on IA and competitive constraints on baryons. When cosmic shear and IGGL are combined, model constraints show an improvement of more than 25\% in $S_8$, also yielding improved and more robust constraints on IA and baryonic feedback, significantly reducing biases in their posteriors. }
\label{fig:iggl_cs}
\end{center}
\end{figure*}

Using a formalism similar to that used in Section \ref{sec:model}, cosmic shear, which correlates observed source ellipticities as described in Eq.~(\ref{eq:ellipticities}), will be independent of galaxy bias and dependent on cosmology and intrinsic alignments. Figure \ref{fig:iggl_cs} shows the model constraints from cosmic shear and realistic IGGL for a selection of the most relevant model parameters in this case: the matter density $\Omega_m$, the amplitude of matter fluctuations $\sigma_8$, the IA amplitude in the NLA model $A_1^\mathrm{IA}$, the baryonic feedback parameter log $T_\mathrm{AGN}$, and the $S_8 \equiv \sigma_8 (\Omega_m /0.3)^{0.5}$ parameter, which measures the best constrained combination of $\Omega_m$ and $\sigma_8$, characteristic of the pure lensing measurements. Again, other varied parameters are not shown as they are either not constrained or dominated by the assumed priors, as described in Table \ref{tab:params}. 

From Fig.~\ref{fig:iggl_cs}, we can see a strong degeneracy between $\sigma_8$ and $\Omega_m$ for both measurements, which we encapsulate in the commonly used $S_8 \equiv \sigma_8 (\Omega_m /0.3)^{0.5}$ parameter, also shown. The IGGL constraints on $S_8$ are weaker than those of cosmic shear, but IGGL yields stronger constraints on IA and competitive constraints on the baryonic feedback model. For this reason, we have also run a combined  MCMC chain from both IGGL and cosmic shear, accounting for the joint covariance between them. The combined IGGL and cosmic shear measurements yield improved constraints in all parameters, and result in an improvement of more than 25\% in $S_8$. In addition, the combination also yields improved and more robust constraints on IA and baryonic feedback, significantly reducing biases in their posteriors, as it can be seen in the figure. 

\subsection{Comparison with clustering cross-correlations}
\label{sec:clusternig}

Finally, there is one other measurement we want to consider for a comparison. As we have seen previously, IGGL is largely dependent on weak lensing magnification, and its cosmological constraining power is largely driven by that effect. In this regard, clustering cross-correlations between galaxy samples that are not overlapping in redshift are also sensitive to weak lensing magnification, and capable of providing cosmological constraints. If we again use a formalism similar to that used in our modeling section (\S\ref{sec:model}), we can get the expression for the clustering cross-correlation of two distinct galaxy samples, 1 and 2: 

\begin{widetext}
\begin{equation}
\label{eq:galaxy_cross}
\begin{split}
 &\left< \delta_g^\text{obs,1} \delta_g^\text{obs,2} \right> = \left< (\delta_g^\text{int,1} +  2 (\alpha_1 -1) \kappa_{l,1}) \,  (\delta_g^\text{int,2} +  2 (\alpha_2 -1) \kappa_{l,2}) \right> = \\
 &= \left< \delta_g^\text{int,1} \delta_g^\text{int,2}\right> 
 + 2 (\alpha_1 -1)\left< \kappa_{l,1} \, \delta_g^\text{int,2} \right> 
 + 2 (\alpha_2 -1)\left< \kappa_{l,2} \, \delta_g^\text{int,1} \right> 
 + 4 (\alpha_1 -1) (\alpha_2 -1)\left< \kappa_{l,1} \, \kappa_{l,2} \right> \simeq 2 b_1 (\alpha_2 -1)\left< \kappa_{l,2} \, \delta_m^\text{1} \right>.
 \vspace{2mm}
\end{split}
\end{equation}
\end{widetext}

We have made several approximations here in arriving at this result. First, we assumed that, if the samples are separated in redshift, then the correlation between intrinsic galaxy densities vanishes. Similarly, the correlation between the intrinsic galaxy density of sample 2 and the convergence of sample 1 do not overlap in redshift, and therefore it vanishes as well. Then, we can assume that the convergence auto-correlation will be smaller than the correlation between intrinsic galaxy density of sample 1 and the convergence of sample 2, so the measurement will be dominated by the latter. In this way, it will depend on the galaxy bias of the first sample, and the magnification coefficient of the second sample. This means that one can use the combination of the clustering auto-correlation of the first sample and the cross-correlation between the two samples to constrain cosmology, in a similar way as it is done for the combination of galaxy clustering and traditional GGL \citep{y3-2x2ptaltlensresults,y3-2x2ptbiasmodelling,y3-3x2ptkp}. 

\begin{figure}
\begin{center}
\includegraphics[width=0.48\textwidth]{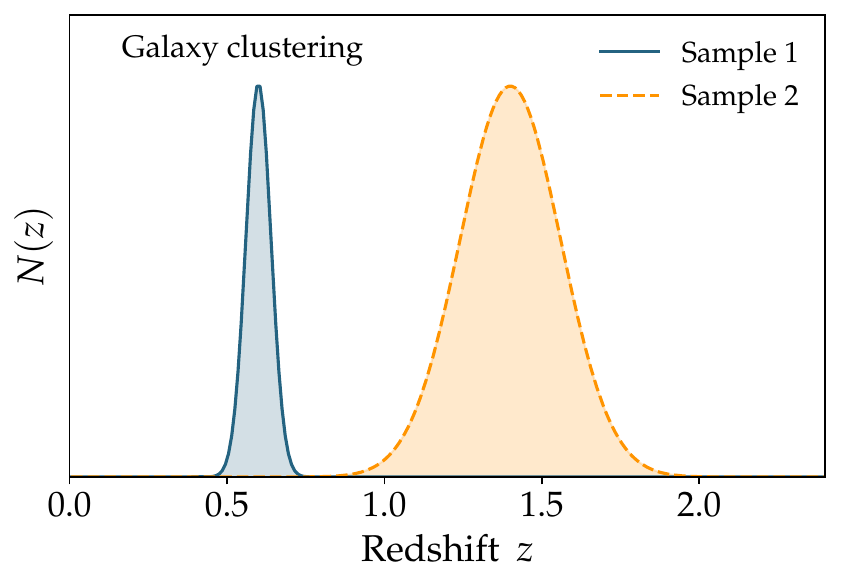}
\caption{Redshift distributions for the two galaxy samples used for the clustering cross-correlation example of Section \ref{sec:clusternig}. As required in order to get weak lensing magnification from galaxy cross-correlations, the redshift distribution have no significant overlap in redshift. Also, the high-redshift bin has the same redshift distribution and galaxy density than that used in the realistic IGGL case, depicted in Fig.~\ref{fig:nzs_theory}. }
\label{fig:nzs_xcorr}
\end{center}
\end{figure}

In order to test this scenario, and compare with IGGL and cosmic shear, we have produced simulated galaxy clustering measurements and covariances in analogy to the GGL and cosmic shear cases studied before. For the two galaxy samples involved in this case, we have used the redshift distributions depicted in Fig.~\ref{fig:nzs_xcorr}. The high redshift sample (sample 2) coincides with the lens (position) sample used in the realistic IGGL case (see bottom panel of Fig.~\ref{fig:nzs_theory}), also in terms of galaxy density and galaxy bias, and the low redshift sample (sample 1) consists of a narrow redshift bin at $z=0.6$, with a galaxy density of 0.3 arcmin$^{-2}$ and galaxy bias $b = 1.4$ , magnification coeffiecient of $\alpha^\mathrm{mag} = 1.25$. 

\begin{figure}
\begin{center}
\includegraphics[width=0.48\textwidth]{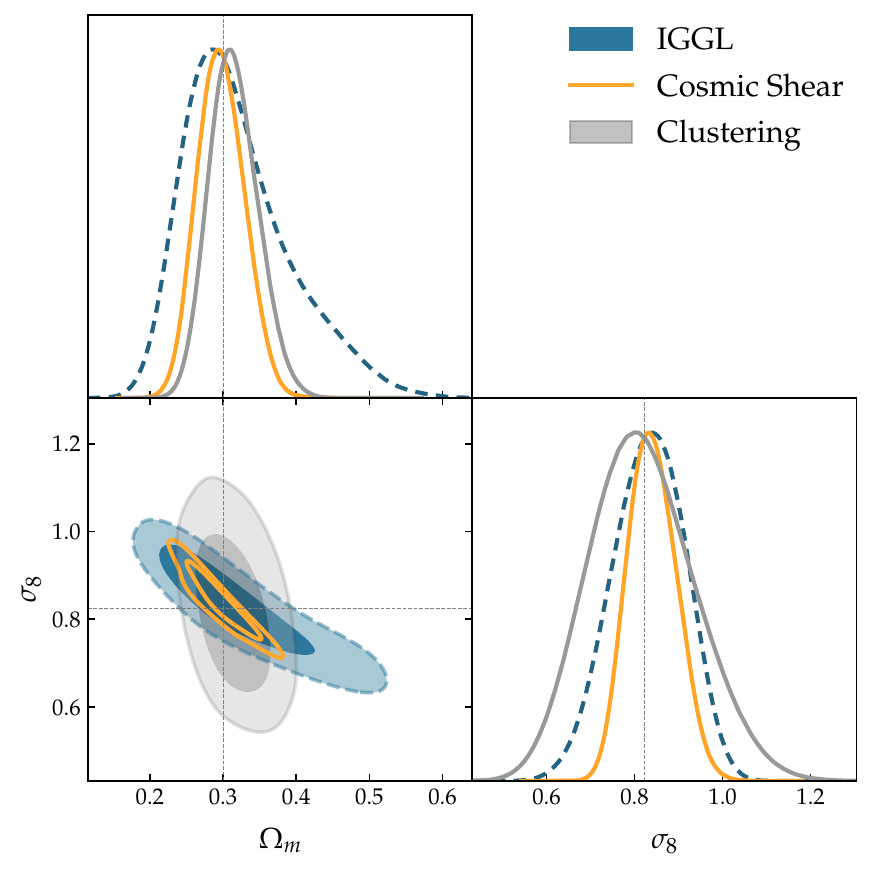}
\caption{Cosmological constraints from IGGL, cosmic shear and clustering cross-correlations as described in Section \ref{sec:model-constraints}. We see how clustering cross-correlations can place strong constraints on $\Omega_m$, but cannot be competitive with cosmic shear and IGGL on $\sigma_8$.}  
\label{fig:constraints_all}
\end{center}
\end{figure}

For the model constraints in this part, we run MCMCs using the galaxy clustering auto-correlation of sample 1 and the galaxy clustering cross-correlation of sample 1 with sample 2 (we use parameter values and ranges as described in Table \ref{tab:params}, and a 5\% Gaussian prior on the magnification coefficient of sample 1). Since galaxy bias is involved in the modeling of both expressions, as shown in Eq.~(\ref{eq:galaxy_cross}), we restrict the analysis to scales larger than 8 $h^{-1}$Mpc for both measurements, for the same reason as the traditional GGL case. Following the GGL case, we fix the log $T_\mathrm{AGN}$ parameter for the clustering MCMCs. Figure \ref{fig:constraints_all} shows the cosmological model constraints from these galaxy clustering measurements, in comparison with the IGGL and cosmic shear constraints. We can see how the clustering measurements can place competitive constrains on $\Omega_m$, but cannot compete with cosmic shear or IGGL in terms of $\sigma_8$ constraining power. 

\section{Discussion} \label{sec:discussion}

The combination of depth and wavelength coverage of upcoming imaging galaxy surveys like the Euclid Satellite and the Vera Rubin Observatory LSST will produce galaxy samples at higher redshifts, compared to the current generation of imaging surveys \citep{Rhodes_2017}. This push to higher redshifts is likely to be more pronounced for lens galaxies, since we only measure positions for those, while source galaxies require the measurement of both positions and shapes (which require more S/N) \citep{highz}. 

For the case of Rubin LSST, the Dark Energy Science Collaboration (DESC) Science Requirements Document (\hyperlink{https://github.com/cosmolike/desc_srd}{SRD}) once stated: `the lens-source bin combinations are only included in the data vector if the lens bin is at lower redshift than the sources (we allow for the lens bin to overlap at 10\%)'. This requirement no longer appears in newer versions of the DESC SRD \citep{thelsstdarkenergysciencecollaboration2021lsstdarkenergyscience}, although it has been used in other publications \citep{Leonard_2024}.  Regardless, that requirement showcases the traditional approach to galaxy-galaxy lensing, in which lens (position) galaxies are only relevant at lower redshifts than source (shape) galaxies. In this paper we have explored a measurement that would not exist if we followed to those requirements.

In previous sections of this paper, we have demonstrated how the measurement of galaxy-galaxy lensing when lens (position) galaxies are at higher redshift than source (shape) galaxies, which we call inverse galaxy-galaxy lensing (IGGL), can provide robust and valuable information to Rubin LSST cosmological analyses. We have proven how the measurement of IGGL is sensitive to weak lensing magnification, intrinsic alignments and cosmology, while being independent of galaxy bias. 

In most cases, IGGL will be driven by weak lensing magnification, and then cosmological information (especially the amplitude of matter fluctuations, $\sigma_8$) will be degenerate with magnification coefficients, which depend on the specific lens (position) galaxy sample \citep{y3-2x2ptmagnification}. This situation is comparable to that of traditional GGL, where cosmology is degenerate with galaxy bias, but with one key difference: galaxy bias cannot be measured reliably from first principles or from simulations, and therefore the combination with galaxy clustering is needed to break the degeneracy and constrain cosmology (which poses several important challenges, see \citep{Noah_weights,Pandey2023}). For the case of IGGL, magnification coefficients can be measured directly from the data, using techniques like source injection \citep{Suchyta2016}, which makes IGGL directly sensitive to cosmology without the need for combination with galaxy clustering measurements.  

Beyond their importance for IGGL, magnification coefficients will become increasingly relevant in cosmological analyses for Rubin LSST and Euclid as their redshift distributions reach higher redshifts, not only for IGGL or traditional galaxy-galaxy lensing but also for galaxy clustering and clustering cross-correlations \citep{Duncan_2013}. Recent analyses show that neglecting the impact of magnification on cosmological analyses combining galaxy clustering, GGL and cosmic shear in Euclid could bias the results by up to 6$\sigma$ \citep{euclid_mag}. In this way, IGGL can also provide a clean way of measuring and validating magnification coefficients, in comparison to the measurements using source injections, and that will be important for the robustness of the cosmological analyses of next-generation surveys. 

On the topic of clustering cross-correlations, which can also be used to get cosmology through weak lensing magnification \citep{Morrison_2012}, we have provided a comparison to IGGL in terms of constraining power. In that case, one key advantage of IGGL is the fact that it does not depend on galaxy bias, and therefore the analyses can use smaller scales provided they also account for baryonic feedback. Furthermore, measuring magnification through clustering cross-correlations requires very little or no redshift overlap between the two tomographic bins (otherwise the measurement is dominated by intrinsic clustering, not magnification) \citep{Morrison_2012}. This is very complicated, especially if one of them is at high redshift and the other needs to be at the peak of the convergence kernel for the high-redshift bin (it would not be efficient to have one at very high redshift and another at very low redshift). This is not a problem for IGGL because, even with overlap between lenses and sources, the overlapping region has very low lensing efficiency and hence the measurement will still be dominated by magnification (as in the realistic IGGL case, not very different from the ideal IGGL case). These reasons make IGGL a very clean and effective way to constrain cosmology from weak lensing magnification, without the need for very precise photometric redshifts.  

Other than magnification, IGGL will also constrain other relevant astrophysical effects. On the one hand, we have shown IGGL to be sensitive to intrinsic alignments, and the IA constraints at high redshift that would be provided by IGGL will be especially relevant for cosmic shear analyses of Rubin LSST and Euclid. On the other hand, IGGL is also sensitive to baryonic effects at small scales, in a way that is not entwined with galaxy bias. These dependencies of IGGL with IA and baryons are important because they can be used to break degeneracies between astrophysics and cosmology in cosmic shear \citep{Leonard_2024}. To that effect, we have showcased how IGGL can improve LSST Year 1 $S_8$ constraints by 25\% while also obtaining improved constraints on IA and baryons. 

Finally, in this paper we have only considered high-redshift lens (position) samples that could be obtained from standard galaxy selections using Rubin LSST data, but we have also shown that IGGL gets more efficient as both lenses and sources move to higher redshifts. In this way, future work should explore the use of other high-redshift galaxy samples as lens (position) samples for IGGL, still using source (shape) galaxy samples from LSST or Euclid. Other high-redshift samples to be used in IGGL can include dropout Lyman-break galaxies \citep{Ono_2017}, high-$z$ sub-millimetre galaxies \citep{Bonavera_2021,bonavera_2024}, high-redshift radio galaxies \citep{Singh_2014}, and quasars \citep{Lyke_2020,juneau2024identifyingquasarsdesibright}. For each different case, IGGL measurements could be used to study the magnification properties of the samples or to constrain cosmology if those properties are already characterized. 







\section{Summary and Conclusions}
\label{sec:conclusions}

Upcoming imaging surveys like the Vera Rubin LSST and the Euclid satellite will yield an order-of-magnitude increase in galaxy sample sizes, which will in turn push these galaxy samples to higher redshifts \citep{Rhodes_2017}. These experiments will also produce weak gravitational lensing measurements with unprecedented precision. The three main observables at the 2-point level are those of galaxy clustering (position-position correlations), galaxy-galaxy lensing (position-shape) and cosmic shear (shape-shape) \citep{y3-3x2ptkp}. For galaxy-galaxy lensing (GGL), lens (position) samples are traditionally at lower redshifts than source (shape) samples, as lens galaxies are tracing the matter field producing shape (shear) distortions on source galaxies \citep{Clampitt2016,Y1GGL,Prat_2022}. In this paper, we explore for the first time the measurement of galaxy-galaxy lensing when lens (position) galaxies are at higher redshift than source (shape) galaxies, which we call inverse galaxy-galaxy lensing (IGGL). 

 Using simulated measurements mimicking an LSST Year 1 survey configuration, we demonstrate how IGGL is sensitive to weak lensing magnification, intrinsic alignments and cosmology, while being independent of galaxy bias. This behavior is very different from traditional GGL, which is heavily sensitive to galaxy bias and therefore needs the combination with galaxy clustering to constrain cosmology \citep{Kwan2016,y3-2x2maglimforecast,y3-2x2ptaltlenssompz,y3-2x2ptbiasmodelling}. This is because traditional GGL is dominated by the correlation between intrinsic galaxy density and shear \citep{Prat_2022}, and that correlation vanishes for IGGL, which is then dominated by weak lensing magnification and IA, with the cosmological dependence coming from the magnification part. At small scales, both traditional GGL and IGGL have dependence on baryonic feedback, but traditional GGL at those scales has intertwined effects from non-linear galaxy bias and baryons, while IGGL is not affected by (non-linear) galaxy bias. This makes IGGL useful to constrain baryonic effects at small scales, without the need for a non-linear bias model. 

We explore the efficiency of IGGL as a function of lens and source redshifts, and find that higher redshifts for both lenses and sources increase the signal-to-noise of the measurement, as the two lensing kernels of magnification and shear increase with redshift. This is equivalent to the redshift trend of cosmic shear, which is also a lensing-lensing correlation. By exploring this efficiency, we are able to differentiate IGGL from traditional GGL, for which signal-to-noise peaks when lens galaxies are about halfway to source galaxies. Also, because of the way magnification and IA affect the galaxy-galaxy lensing observable, IGGL can yield negative signals when measured around high-redshift galaxies. This is also a difference with respect to traditional GGL, which can only be negative around cosmic voids \citep{Krause_2012,Melchior_2014,2017MNRAS.465..746S}, not galaxies. 

Using Monte Carlo Markov Chains (MCMC) and a full cosmological and nuisance parameter model, we determine the constraining power of IGGL, and compare it to traditional GGL. Because IGGL does not depend on galaxy bias, we can use smaller scales as long as we allow for baryonic feedback to vary in our model, which we do through the $T_\mathrm{AGN}$ parameter \citep{HMCODE}. Even if IGGL measurements have a smaller signal-to-noise, we show how IGGL outperforms traditional GGL in measurements of $\sigma_8$ and $A^\mathrm{IA}_1$, which are the amplitudes of matter fluctuations and tidal alignment in our IA model, while also providing a constraint on $T_\mathrm{AGN}$. IGGL does not depend on galaxy bias and therefore does not show a degeneracy between $b$ and $\sigma_8$, like traditional GGL does. Instead, it shows a degeneracy between $\sigma_8$ and $\Omega_m$, sometimes summarized in the $S_8 \equiv \sigma_8 (\Omega_m /0.3)^{0.5}$ parameter, characteristic of the pure lensing measurements like cosmic shear. 

Examining how to maximize the constraining power of IGGL, we found that better priors on magnification and IA strengthen our posteriors on $S_8$. These improved priors are on the horizon, with models like NLA and TATT being applied to multiple weak lensing datasets \citep{Blazek_2019, Blazek_2015, Blazek2012, blazek2011, Zhang_2010, DESy1, Hildebrandt2018, asgari_2021}, and multiple approaches for measuring magnification being developed and tested \citep{y3-2x2ptmagnification, bonavera_2024, Euclidmagnification2021, ma_2024, Wenzl_2024, Qin_2023}.

When comparing IGGL constraints to cosmic shear, which has a much larger signal-to-noise, we find IGGL gives stronger constraints on IA and competitive constraints on baryonic feedback, while cosmic shear gives stronger constraints on $S_8$. When we combine IGGL and cosmic shear, taking into account the joint covariance, the precision of the $S_8$ measurement increases by more than 25\% with respect to cosmic shear alone, while constraints on IA and baryons also improve significantly. Finally, we compare IGGL to measurements of clustering cross-correlations of non-overlapping redshift bins, which are also a probe of magnification. We find clustering cross-correlations to yield strong constraints on $\Omega_m$, but IGGL provides much tighter constraints on $\sigma_8$. 

After all the analysis and findings in this paper, we conclude that IGGL, a measurement that has not been considered before or even been discarded, will provide important constraints on cosmology and also astrophysical effects like intrinsic alignments and baryonic feedback, while providing independent constraints on magnification parameters that will be key to get unbiased cosmology from upcoming surveys \citep{euclid_mag}. The combination with cosmic shear will significantly improve the precision of the $S_8$ measurement while also adding robustness to the analysis in terms of IA and baryons. 

Of the publicly available galaxy datasets made for cosmology, most of the redshift bins that have been released contain galaxies of redshift of z $\lesssim$ 1.5 (e.g. \citep{DESy1, y3-sompz, speagle2019}). As we have shown, IGGL becomes most effective with lens (position) redshifts higher than that, and so these weak-lensing-specific datasets are not at high enough redshifts to give effective IGGL measurements. Recently, a high-redshift galaxy catalog from the Dark Energy Survey has been released \citep{highz}. While this dataset is still not ideal for IGGL, it can still serve as an initial testing ground for upcoming surveys that will reach higher redshifts. Finally, beyond the use of lens samples that were made specially for weak lensing cosmology, measurements of IGGL can also be explored with other high-redshift samples such as those of Lyman-break galaxies \citep{Ono_2017}, quasars \citep{Lyke_2020,juneau2024identifyingquasarsdesibright}, sub-millimetre \citep{Bonavera_2021,bonavera_2024} and radio galaxies \citep{Singh_2014}.

\begin{acknowledgments}
The authors thank Alex Alarcon, Santi Avila, Eric Baxter, Bhuvnesh Jain, Ramon Miquel and Judit Prat for
helpful conversations about this topic. DNC acknowledges support from the predoctoral program AGAUR Joan Oró of the Secretariat of Universities and Research of the Department of Research and Universities of the Generalitat of Catalonia and the European Social Plus Fund (2023 FI-1-00683). CS is supported by  Ramón y Cajal project RYC2021-031194-I, funded by MCIN/AEI/
10.13039/501100011033 and by the “European Union
NextGenerationEU/PRTR”. This work was supported by project PID2022-138123NA-I00, funded by MCIN/AEI/
10.13039/501100011033. 
\end{acknowledgments}



\bibliographystyle{JHEP} 
\bibliography{library,apssamp}

\end{document}